\documentclass[showpacs,superscriptaddress,floatfix,amssymb,amsmath,eqsecnum,aps,nofootinbib]{revtex4}

\usepackage{slashed}
\usepackage{graphicx}
\usepackage{verbatim}
\usepackage{amsthm}
\usepackage{amsfonts}
\usepackage{amscd}
\usepackage{epsfig}
\usepackage[dvips]{color} 

\def\3by3Mat#1#2#3#4#5#6#7#8#9{\ensuremath{\begin{pmatrix}#1&#2&#3\\
                                                          #4&#5&#6\\
                                                          #7&#8&#9\end{pmatrix}}}

\def\fivecMat#1#2#3#4#5{\ensuremath{\begin{pmatrix}#1\\#2\\#3\\#4\\#5\end{pmatrix}}}
\def\threecMat#1#2#3{\ensuremath{\begin{pmatrix}#1\\#2\\#3\end{pmatrix}}}

\def\twocMat#1#2{\ensuremath{\begin{pmatrix}#1\\#2\end{pmatrix}}}

\def\msol{\ensuremath{m_\textrm{sol}}}
\def\matm{\ensuremath{m_\textrm{atm}}}
\def\order#1{\ensuremath{\mathcal{O}\left(#1\right)}}
\def\vev#1{\ensuremath{\langle #1\rangle}}
\def\mnu{\ensuremath{m_\nu}}
\def\mnud{\ensuremath{m_\nu^D}}
\def\mnuhat{\ensuremath{\widehat{m}_\nu}}
\def\Upmns{\ensuremath{U_\text{PMNS}}}
\def\Uckm{\ensuremath{U_\text{CKM}}}

\def\Uw{\ensuremath{U_\omega}}
\def\MatU{\ensuremath{\frac{1}{\sqrt{3}}
                         \begin{pmatrix}1 &1 &1\\
                                     1 &\omega &\omega^2\\
                                     1 &\omega^2 &\omega\end{pmatrix}}}
\def\Matdiag#1#2#3{\ensuremath{\begin{pmatrix}#1&0&0\\
                                              0&#2&0\\
                                              0&0&#3\end{pmatrix}}}
\def\rept{\ensuremath{\underline{3}}}
\def\rep{\ensuremath{\underline{1}}}
\def\repp{\ensuremath{\underline{1}'}}
\def\reppp{\ensuremath{\underline{1}''}}

\begin{document}

\title{Determining the heavy seesaw neutrino mass matrix from low-energy parameters}

\author{Xiao-Gang He}\email{hexg@phys.ntu.edu.tw}
\affiliation{Department of Physics, Center for Theoretical Sciences and LeCosPA Center, National Taiwan University, Taipei 10617, Taiwan, Republic of China}
\author{Sandy S. C. Law}\email{slaw@ph.unimelb.edu.au}
\affiliation{School of Physics, The University of Melbourne, Victoria 3010, Australia}
\author{Raymond R. Volkas}\email{raymondv@unimelb.edu.au}
\affiliation{School of Physics, The University of Melbourne, Victoria 3010, Australia}

\date{October 7, 2008}

\begin{abstract}
We explore how the seesaw sector in neutrino mass models may be constrained through symmetries to be completely determined
in terms of low-energy mass, mixing angle and $CP$-violating phase observables.  The key ingredients are intra-family symmetries to
determine the neutrino Dirac mass matrix in terms of the charged-lepton or quark mass matrices, together with inter-family or
flavor symmetries to determine diagonalization matrices.  Implications for leptogenesis and collider detection of heavy neutral
leptons are discussed.  We show that leptogenesis can succeed in small regions of parameter space for the case where the neutrino
Dirac mass matrix equals the up-quark mass matrix.  The model where the neutrino Dirac mass matrix equals the charged-lepton mass matrix
can yield a heavy neutral lepton as light as about 1 TeV, but detecting such a particle will be difficult.
 
\end{abstract}


\maketitle

\section{Introduction}

Neutrino oscillation experiments involving neutrinos and antineutrinos coming from astrophysical and terrestrial sources \cite{neutrinos_exp} have found compelling evidence that neutrinos have mass. To accommodate this observation, the minimal Standard Model (SM) must be extended. Some sensible  ways  to do this include: 
(a) Type I seesaw with three heavy right-handed (RH) Majorana neutrinos \cite{type1_seesaw}, 
(b) the use of an electroweak Higgs triplet to directly provide the left-handed (LH)
neutrinos with small Majorana masses (Type II seesaw \cite{type2_seesaw}),
(c) introducing a fermion triplet (Type III seesaw \cite{type3_seesaw}),
(d) the generation of three Dirac neutrinos through an exact parallel of
the SM method of giving mass to charged fermions,  and (e) the radiative generation of neutrino masses as per the Zee or Babu models \cite{zeebabu}. 
But in the absence of more experimental data, it is impossible to tell which, if any, of these is actually correct. 

The focus of this paper is on method (a), the seesaw framework with three heavy RH Majorana neutrinos (denoted $N$ throughout). It is an attractive possibility because it simply posits the existence of these $N$'s to parallel the multiplet structure of the other fermions while providing a simple
explanation for why the light neutrinos are so much less massive than the charged leptons. Since the setup uses the most general renormalisable Lagrangian consistent with the SM gauge symmetry, both the Yukawa couplings of the LH leptons to the RH neutrinos and bare Majorana masses are permitted for the RH neutrinos. Consequently, the additional assumption that the RH Majorana mass scale is much higher than that of the charged fermions leads to a tiny mass for ordinary neutrinos through the famous seesaw relation:
$\mnu \sim m_f^2/M_R$, where $\mnu$ is the Majorana mass for a light neutrino and $M_R$ is a large RH Majorana mass ($M_R \gg m_f$) with $m_f$ being most naturally of the order of a charged fermion mass. The three light neutrino mass eigenstates are accompanied by three heavy neutral lepton mass
eigenstates.

Depending on the parameter space for the RH neutrino bare masses and Yukawa couplings, additional benefits may flow: thermal
leptogenesis \cite{Fukugita:1986hr} if there are appropriate $CP$-violating decays and if the lightest of the heavy $N$'s has mass greater than about $10^{9}$ GeV  \cite{BDP}; 
leptogenesis through $CP$-violating oscillations of the $N$'s as in the Akhmedov-Rubakov-Smirnov mechanism \cite{ARS}; or $N$'s as a warm dark matter candidate \cite{WDM} \footnote{Note that due to the constraints in the parameter space, this scenario cannot really be called a seesaw model, but the form of the Lagrangian is the same.}.

Since the mass eigenstate heavy neutral leptons are to a good approximation sterile with respect to
gauge interactions, they are difficult to detect experimentally.  
This is especially true if they are also extremely massive, as in the thermal leptogenesis alternative.  On the other hand, if they are not as 
massive and are in the TeV scale, then they can be looked for in colliders through their Yukawa interactions,
and through their suppressed but nonzero weak interactions (induced through the mass mixing with regular active neutrinos).

To experimentally test the seesaw scenario, it would be helpful if one knew the parameters governing the $N$-sector including their interactions
with other SM particles.  In the minimal seesaw model, these parameters are arbitrary, so one has to go beyond the minimal model to achieve this goal.
The purpose of this paper is to illustrate how symmetries may be used to determine the RH Majorana mass matrix as a function of low-energy mass, mixing angle and $CP$-violating phase observables by constructing several representative models. We then examine these models to see if thermal leptogenesis can succeed or if experimentally accessible heavy $N$'s are predicted.

In the next Section we discuss the general model building symmetry requirements for relating the $N$-sector parameters to 
low-energy observables.  Section \ref{Sec:symmetry-basics} then revises the basic properties of the identified symmetries, followed
by Sec.~\ref{Sec:models} which details specific models.  Section \ref{Sec:phenomenology} presents a phenomenological study of those
models, and we conclude in Sec.~\ref{Sec:conclusion}

\section{Seesaw structure and relation to the low-energy sector}
\label{Sec:low-high}

The effective light Majorana neutrino mass matrix $\mnu$, defined through
\begin{equation}
\frac{1}{2}\, \overline{\nu} \; \mnu \, \nu^c + \text{h.c.} \;,
\end{equation}
is given by
\begin{equation}
\mnu = -\mnud\, M_R^{-1} (\mnud)^T + \order{(\mnud)^3/M_R^2} \;,
\label{eq:ML}
\end{equation}
where $\mnud$ is the neutrino Dirac mass matrix, defined through
\begin{equation}
 \overline{\nu}_L \,\mnud\, \nu_R + \text{h.c.} \;,
\end{equation}
whilst the RH Majorana mass matrix $M_R$ is defined through
\begin{equation}
\frac{1}{2} \,\overline{(\nu_R)^c}\, M_R\, \nu_R + \text{h.c.} \;.
\end{equation}

Let
\begin{equation}
\nu^m = V_{\nu} \;\nu
\label{eq:numassbasis}
\end{equation}
be the mass-eigenstates for the light Majorana neutrinos, where $V_{\nu}$ is the unitary diagonalization matrix for $\mnu$. The diagonalized $\mnu$ is therefore \footnote{Diagonalized matrices will always be denoted by a carat in this paper.}
\begin{equation}
\mnuhat \equiv {\rm diag}(m_1, m_2, m_3) = -V_{\nu} \,\mnu\, V^{T}_{\nu}\;.
\label{eq:MLdiag}
\end{equation}
Eq.(\ref{eq:ML}) then implies that
\begin{equation}
\mnuhat \simeq  V_{\nu}\, \mnud \,M_R^{-1} (\mnud)^T V^{T}_{\nu}\;.
\label{eq:ML2}
\end{equation}
The matrix $\mnuhat$ has been experimentally determined up to an absolute light neutrino mass scale, which we shall
conveniently parameterize through the lightest $\mnu$ eigenvalue. For the normal hierarchy (NH) with $m_1 < m_2 < m_3$, we have (for $m_{1,2,3} \in \mathbb{R}^+ \cup \{0\}$)
\begin{equation}
m_2 = \sqrt{m_1^2 + \Delta \msol^2}
\qquad,\qquad
m_3 = \sqrt{m_1^2 + \Delta \msol^2 + \Delta \matm^2} \;\;,
\label{eq:NH}
\end{equation}
where $\Delta\msol^2 \simeq 7.7 \times 10^{-5}$ eV$^2$ and $\Delta\matm^2 \simeq 2.4 \times 10^{-3}$ eV$^2$ are the ``solar'' and ``atmospheric'' squared-mass difference respectively \cite{neutrinos_exp, mixingdata, Schwetz:2008er}.
For the inverted hierarchy (IH) with $m_3 < m_1 < m_2$, we obtain
\begin{equation}
m_1 = \sqrt{m_3^2 + \Delta \matm^2 - \Delta \msol^2}
\qquad,\qquad
m_2 = \sqrt{m_3^2 + \Delta \matm^2} \;\;.
\label{eq:IH}
\end{equation}
In order to connect the high- and low-energy sectors, one must have $M_R$ completely determined by known quantities. Hence, our goal is to have $M_R$ constructed from some combination of $\mnuhat$, the charged-fermion mass matrices, $\widehat{m}_f$ with $f=e,d,u$, and the lepton and quark mixing matrices ($\Upmns$ and $\Uckm$) respectively. As a consequence, the first necessary
condition, according to Eq.(\ref{eq:ML2}), is:
\begin{equation}
\text{\em The neutrino Dirac mass matrix, \mnud, must be predicted by the theory.}  
\label{eq:cond1}
\end{equation}
The simplest possibility is that
\begin{equation}
\mnud = m_f\ \ {\rm for\ one\ of}\ f=e,\ f=d\ {\rm or}\ f=u\;.
\label{eq:massrelation}
\end{equation}
There are custodial $SU(2)$, unification and quark-lepton symmetries that can enforce each of these conditions at tree-level, 
as we shall review in the next section.  For the moment, let us just accept that they are all possible. Equation (\ref{eq:ML2}) now becomes
\begin{equation}
\mnuhat \simeq  V_{\nu}\, m_{f}\, M_R^{-1}\, m_{f}^T\, V^{T}_{\nu}\;.
\label{eq:ML3}
\end{equation}
Introducing the diagonalized fermion mass matrix
\begin{equation}
\widehat{m}_f = V_{fL}\, m_f\, V_{fR}^{\dagger}\;,
\label{eq:hatmf}
\end{equation}
where the $V_{fL}$ and $V_{fR}$ are the left- and right-diagonalization matrices for $m_f$ respectively, Eq.(\ref{eq:ML3}) can be rewritten as
\begin{align}
\mnuhat &\simeq 
 V_{\nu}\, V_{fL}^{\dagger}\, \widehat{m}_{f}\, V_{fR}\, M_R^{-1}\, V_{fR}^{T}\, \widehat{m}_{f}\, V_{fL}^*\, V^{T}_{\nu}\;,\nonumber\\
 &=
 (V_{fL}\,V_{\nu}^{\dagger})^{\dagger}\, \widehat{m}_{f}\, V_{fR}\, M_R^{-1}\, V_{fR}^{T}\, \widehat{m}_{f}\, (V_{fL}\,V_{\nu}^{\dagger})^{*}\;,
\label{eq:ML4}
\end{align}
which in turn reveals the second necessary condition: 
\begin{equation}
\text{\em One has to know the diagonalization matrix product: } V_{fL} V^{\dagger}_{\nu}\; \text{\em and the right-diagonalization matrix}, V_{fR}.
\label{eq:cond2}
\end{equation}
Because the known weak interactions are left-handed, the right-diagonalization matrix cannot be measured.\footnote{Of course, the discovery of
right-handed weak interactions would change this situation.}  
Therefore, to satisfy condition (\ref{eq:cond2}), $V_{fR}$ needs to be determined by the theory, and this usually means a \emph{flavor symmetry} is required \footnote{See \cite{Berezhiani:1990jj} for an earlier work on flavor symmetry and the seesaw mechanism.}. 
In the next section we shall review how flavor symmetries can give rise to fully determined diagonalization matrices, where their entries are just numbers, usually related to the Clebsch-Gordan coefficients of the flavor symmetry group.

The product
$V_{fL} V^{\dagger}_{\nu}$
is similar in form to both the PMNS and CKM matrices, which are, respectively,
\begin{equation}
\Upmns = V_{eL} V_{\nu}^{\dagger}
\qquad,\qquad \Uckm = V_{uL} V_{dL}^{\dagger}\;.
\label{eq:PMNSCKMdefns}
\end{equation}
The simplest ansatze are that $V_{fL} V^{\dagger}_{\nu L}$ equals either $\Upmns$ or $\Uckm$.  The next simplest class would
see $V_{fL} V^{\dagger}_{\nu}$ equal to a product of either the PMNS or CKM matrix and a known matrix predicted by the flavor symmetry selected.

Let us consider some special cases.  The simplest possibility suggested by the above equations would be that
\begin{equation}
f = e\quad {\rm and}\quad V_{eR} = 1\;.
\label{eq:e_PMNS}
\end{equation}
These relations may be achieved by imposing a $(\nu \leftrightarrow e)$ and flavor symmetry respectively.  The RH Majorana mass matrix is then
completely determined through
\begin{equation}
M_R \simeq \widehat{m}_e\, \Upmns^* \,\mnuhat^{-1}\, \Upmns^{\dagger}\, \widehat{m}_e \;.
\label{eq:MR_e_PMNS}
\end{equation}
Two other possibilities, arising from the enforcement of ($\nu \leftrightarrow d,u$) and the appropriate flavor symmetries, are that
\begin{eqnarray}
& f = d\;, \qquad\text{  with  }\;\; V_{dR} = 1\quad,\quad V_{dL} = V_{eL}\;, & \label{eq:d_PMNS}\\
\text{and  }
& f = u\;, \qquad\text{  with  }\;\; V_{uR} = 1\quad,\quad V_{uL} = V_{eL}\;,& \label{eq:u_PMNS}
\end{eqnarray}
leading to
\begin{equation}
M_R \simeq  \widehat{m}_{d,u}\, \Upmns^*\, \mnuhat^{-1}\, \Upmns^{\dagger}\, \widehat{m}_{d,u}\;.
\label{eq:MR_ud_PMNS}
\end{equation}

Because of the automatic presence of $V_{\nu}$ in the formula for $M_R$, it is relatively straightforward to find
symmetries leading to Eqs.~(\ref{eq:MR_e_PMNS}) and (\ref{eq:MR_ud_PMNS}) where the leptonic PMNS mixing matrix is
a key feature.  But it may also be of interest to consider symmetry structures that can lead to the PMNS matrix being
replaced by the CKM matrix (or a product of the two).  One way to try this would be to arrange symmetries such that $\mnu$ would be necessarily diagonal,
giving $V_{\nu} = 1$.  Then, the condition $f = d$, together with $V_{uL} = V_{dR} = 1$, will lead to $\Uckm = V_{dL}^{\dagger}$, and hence the relation
\begin{equation}
M_R \simeq  \widehat{m}_{d}\, \Uckm^T\, \mnuhat^{-1}\, \Uckm\, \widehat{m}_{d}\;,
\label{eq:MR_d_CKM}
\end{equation}
would be obtained.  A similar relation with $d$ and $u$ interchanging roles could equally well be contemplated.  The delicate
part would be obtaining a diagonal $\mnu$ without forcing a diagonal $\mnud$.  If the latter
were diagonal, then the relations $\mnud = m_d$ or $\mnud = m_u$ would also imply that $V_{dL} = 1$ or respectively
$V_{uL} = 1$, and hence leading to $\Uckm = 1$ at tree-level.

Finally, there is of course the relatively mundane case where all of the diagonalization matrices in the formula for $M_R$ are equal to the identity, so that one simply gets
\begin{equation}
M_R \equiv \widehat{M}_R  \simeq \text{diag}\left(\frac{m_{f1}^2}{m_1},\frac{m_{f2}^2}{m_2},\frac{m_{f3}^2}{m_3}
\right)\;.
\label{eq:mundane}
\end{equation}
Interestingly, this is not possible for the $f = e$ choice, because the PMNS matrix is known to be very dissimilar to the identity.  However, flavor symmetries allowing, Eq.(\ref{eq:mundane}) can in principle be achieved for $f= d$ or $u$. In these situations, one would then get $\Upmns = V_{eL}$ and $\Uckm = V_{uL}\;(\text{if } f=d)$ and $V_{dL}^\dagger\;(\text{if } f=u)$.

Although the analysis above was framed in terms of the leading seesaw expression 
$\mnu \simeq - \mnud M_R^{-1} (\mnud)^T$, it
generalizes to cases where additional terms on the right-hand side are kept, because the higher-order terms contain {\it a priori} the
same unknowns as does the leading term.

In summary, the general properties of enforcing a ($\nu \leftrightarrow e,d,u$) symmetry in parallel with some flavor symmetries motivate relations of the form
\begin{equation}
M_R = M_R(\widehat{m}_e\,, \widehat{m}_d\,, \widehat{m}_u\,, \Upmns\,, \Uckm)
\label{eq:generalformula}
\end{equation}
of which Eqs.(\ref{eq:MR_e_PMNS}), (\ref{eq:MR_ud_PMNS}), (\ref{eq:MR_d_CKM}) and (\ref{eq:mundane}) are important examples.

\section{The use of Symmetries}
\label{Sec:symmetry-basics}

The aim of this section is to briefly illustrate how mass relations of the type
$\mnud = m_{e,d \text{ or } u}$ may be enforced, as well as the role of flavor symmetry in determining the diagonalization matrices of interest. We will present some concrete examples that utilize these ideas to good effect in the next section.

It is well known that in a minimal $SO(10)$ framework one obtains the mass relations $\mnud = m_e = m_d = m_u$, because all fermions are in the same multiplet and the electroweak Higgs lies in a real fundamental of $SO(10)$.  
These relations are too powerful from a phenomenological perspective: while the neutrino Dirac mass matrix is related to that of 
another fermion as desired, the other mass relations $m_e = m_d = m_u$ are {\it not} wanted.  However, this observation
motivates the search for gauge groups that contain the SM as a subgroup and have enough power to establish the mass relation we seek without violating any known experimental constraints. Indeed, subgroups of $SO(10)$ are good starting points for such a search. Outside of $SO(10)$, the 
use of discrete rather than continuous symmetries to relate different multiplets constitutes another sensible strategy.

Let us consider the following groups, motivated by being subgroups of $SO(10)$, but not necessarily to be thought of as arising from an underlying
$SO(10)$ theory: the standard $SU(5)$ unification group \cite{Georgi:1974sy}, its flipped extension $SU(5) \otimes U(1)$ \cite{Barr:1981qv} and
the Left-Right group $SU(3)_c \otimes SU(2)_L \otimes SU(2)_R \otimes U(1)_{B-L}$  \cite{LR_Mohapatra.Pati}.  Standard $SU(5)$ has the LH charged leptons and LH down antiquarks
in the $\overline{5}$ representation, while their mass partners are in the $10$.  In the minimal model a single Yukawa term couples those two multiplets to
a Higgs in the $\overline{5}$, leading to the relation $m_e = m_d$.  The up-quark and neutrino Dirac masses are governed by independent Yukawa couplings, so
they are unrelated to each other and unrelated to $m_e$ and $m_d$.  In flipped $SU(5)$, the down antiquarks and the up antiquarks flip roles, as also
do the charged antileptons and antineutrinos. The minimal model thus supplies $\mnud = m_u$ with unrelated $m_d$ and $m_e$ entries.  For our purposes,
standard $SU(5)$ is not useful, but flipped $SU(5)$ is interesting\footnote{The Pati-Salam-like \cite{Pati:1974yy} subgroup $SU(4) \otimes SU(2)_L \otimes U(1)_R$ can also be used to enforce $\mnud = m_u$}.  The third subgroup, the Left-Right group, has the power to enforce mass
degeneracy between weak isospin partners: $m_d = m_u$ and $m_e = \mnud$  \cite{Volkas:1995yn}. Such a degeneracy follows from requiring
a bidoublet Higgs to be {\it real}, which at the $SO(10)$ level follows from having the Higgs 10-plet being real.
This basically causes $SU(2)_R$ to become custodial $SU(2)$.  So we conclude that flipped $SU(5)$ which can give $\mnud = m_u$ and
the Left-Right group which can give $\mnud = m_e$ are relevant $SO(10)$ subgroups for our purposes.

The other obvious mass relation $\mnud = m_d$ will be obtained in the next section not from $SO(10)$ or any of its subgroups, but rather
by using the idea of discrete quark-lepton symmetry \cite{QL_symmodel}.  The idea here is to extend the gauge group by including an $SU(3)$ color group for leptons,
with standard leptons identified as one of the colours after spontaneous symmetry breaking.  The gauge structure now permits a discrete
interchange symmetry between quarks and (generalized) leptons to be imposed, from which $\mnud = m_d$ can follow.

Though we shall not pursue this line of thought further in this paper, we should also remark that the relation between 
$\mnud$ and $m_{e,d \text{ or } u}$
need not be a direct equality.  At the $SO(10)$ level one can consider embedding the electroweak Higgs doublet not in the $10$ but in a
higher-dimensional representation.  In that case a matrix of Clebsch-Gordan coefficients relates $\mnud$ with the other fermion mass
matrix, as a generalization of the well-known Georgi-Jarlskog \cite{Georgi:1979df}
 modification of the $m_e$ to $m_d$ relation in $SU(5)$ unification.

Once the appropriate fermion-mass-constraining group is selected, the remaining challenges are twofold. The first, as well-illustrated
by minimal $SO(10)$, is the removal of byproducts such as unwanted mass relations or interactions. The second is the need to have predictable diagonalization matrices. Quite frequently, it is possible to meet both of these challenges by introducing a \emph{flavor symmetry} and a non-minimal Higgs sector. In cases where this is not sufficient, unbroken global non-flavor symmetries may be imposed to eliminate all undesirable terms.

The key concept is that of a ``form-diagonalizable matrix''  \cite{Low:2003dz}.  This is a matrix containing relations amongst its
elements and perhaps also texture zeros so as to make the diagonalization matrices fully determined while leaving the eigenvalues arbitrary.  Special flavor symmetries exist
to enforce form-diagonalizability, and they have in recent years been widely used to try to understand the ``tribimaximal'' form \cite{TBmixing}
that is consistent with the experimentally measured PMNS matrix.

In the models presented below, combined effect of the mass-relating symmetry and the flavor symmetry will be to produce a relation of 
the form $\mnud = K\,\widehat{m}_{e,d \text{ or } u}$, where $K$ is given by a known diagonalization matrix.

\section{Some Representative Models}\label{Sec:models}

In this section, we construct three realistic models that can enforce 
$\mnud = K\,\widehat{m}_{e,d \text{ or } u}$, and subsequently lead to relations (\ref{eq:MR_e_PMNS}) and (\ref{eq:MR_ud_PMNS}) respectively.

\subsection{Relating $\mnud$ to $\widehat{m}_u$ via a flipped $SU(5)$ model}

We consider a flipped $SU(5)$ group \cite{Barr:1981qv} augmented by $A_4$ flavor symmetry \cite{A4_ref, He:2006dk}:
\begin{align}
 G_1 &= SU(5) \otimes U(1)_X \times A_4\;,\label{eq:su5_1}\\
 &\supset SU(3)_c \otimes SU(2)_L \otimes \underbrace{U(1)_T \otimes U(1)_X}_{U(1)_Y} \times \,A_4 \;,\label{eq:su5_2}
\end{align}
with hypercharge $Y$ given by a linear combination of $T$ and $X$. The choice of this gauge group is for the reason discussed in the
previous section: one naturally obtains the useful mass relation $\mnud = m_u$ while avoiding $m_e = m_d$. The role of the flavor symmetry is then purely to ensure that all diagonalization matrices are completely determined.

For this model, the particle contents and their transformation properties under $G_1$ are given by:
\begin{align}
 \psi_{L\alpha} &= \fivecMat{u_{R}^{1c}}{u_{R}^{2c}}{u_{R}^{3c}}{e_L}{-\nu_L}
                  \;\sim\; (\overline{5}, -3)(\rept)
\;\; ; \;\;
 \chi_L^{\alpha\beta} = \frac{1}{\sqrt{2}}
  \begin{pmatrix}
    0       & d_{R}^{3c}  & -d_{R}^{2c}  & -u_{L}^1 & -d_{L}^1 \\
  -d_{R}^{3c} &    0      & d_{R}^{1c}   & -u_{L}^2 & -d_{L}^2 \\
   d_{R}^{2c} & -d_{R}^{1c} &    0       & -u_{L}^3 & -d_{L}^3 \\
   u_{L}^1   & u_{L}^2    & u_{L}^3     &    0    & -\nu_R^c\\
   d_{L}^1   & d_{L}^2    & d_{L}^3     & \nu_R^c &    0    \\
  \end{pmatrix}
  \;\sim\; (10,1)(\rep \oplus\repp\oplus\reppp) \;; \nonumber\\
  e_R^c &\sim\; (1,5)(\rep \oplus\repp\oplus\reppp) \;\; ; \;\;
  \Phi_{(3)}^\sigma = \fivecMat{h_{1d}}{h_{2d}}{h_{3d}}{\phi_{(3)}^{0*}}{-\phi_{(3)}^{+}}
  \;\sim\;(5,-2)(\rept)\;\; ;\;\;
  \Phi_{(1\oplus 1' \oplus 1'')}^\sigma \;\sim\;(5,-2)(\rep \oplus\repp\oplus\reppp)\;\;;\nonumber\\
  &\Delta^{\alpha\beta\gamma\delta} \;\sim\; (\overline{50},2)(\rep \oplus\repp\oplus\reppp)\;,
  \label{eq:SU5_content}
\end{align}
where the superscripts 1,2 and 3 and Greek letters are the color and $SU(5)$ indices respectively. In matrix form, the $G_1$ invariant interaction Lagrangian then contains the following terms:
\begin{align}
 -\mathcal{L} &=
    Y_{\lambda 1}\, \overline{\psi}_L\, \Phi_{(3)}^*\, e_R
  + \sqrt{2}\;Y_{\lambda 2}\; \overline{\psi}_L\, \chi_L^c \,\Phi_{(3)}
  + \frac{Y_{\lambda 3}}{4}\, (\overline{\chi}_L)_{\alpha\beta} (\chi_L^c)_{\gamma\delta}
   \left(\Phi^*_{(1\oplus 1'\oplus 1'')}\right)_\sigma
    \,\epsilon^{\alpha\beta\gamma\delta\sigma} \nonumber\\
  &\qquad
  + Y_{\lambda 4}  (\overline{\chi}_L)_{\alpha\beta} (\chi_L^c)_{\gamma\delta}
    \,\Delta^{\alpha\beta\gamma\delta} + \text{h.c.}\;,
    \label{eq:SU5_Lag1}
\intertext{and when the neutral components of $\Phi$ and $\Delta$ obtain nonzero VEVs, one gets mass terms of the form}
 &= Y_{\lambda 1}\, \overline{e}_L\, \vev{\phi^0_{(3)}}\, e_R
   - Y_{\lambda 2}\, (
   \overline{u}_L\,\vev{\phi^{0*}_{(3)}}\,u_R
    +
      \overline{\nu}_L\,\vev{\phi^{0*}_{(3)}}\,\nu_R)
   + \frac{Y_{\lambda 3}}{2}\, \left(\overline{d^c_R}\,d^c_L + \overline{d}_L\,d_R \right)\vev{\phi^0_{(1\oplus 1'\oplus 1'')}} \nonumber\\
 &\qquad
   + Y_{\lambda 4}\,\overline{\nu_R^c}\,\nu_R \vev{\Delta^0_{(1\oplus 1'\oplus 1'')}} + \text{h.c.}\;.
   \label{eq:SU5_Lag2}
\end{align}
Note that $\vev{\Delta}$, which provides the heavy Majorana mass, breaks $G_1$ down to the SM, and is expected to be at a much higher energy scale than $\vev{\Phi}$ which breaks electroweak symmetry.

Writing out the $A_4$ structure of the $Y_{\lambda 1}$- and $Y_{\lambda 2}$-terms in Eq.~(\ref{eq:SU5_Lag2}) with the vacuum $\vev{\phi_{(3)}^0} \equiv \vev{\phi_{(3)}^{0*}}=  (v_{(3)},v_{(3)},v_{(3)})$ where $v_{(3)} \in \mathbb{R}$, one gets
\begin{align}
 m_e:\,\quad
 & \lambda_1\, (\overline{e}_L\,\vev{\phi_{(3)}^0})_{\underline{1}}\,e_R
 + \lambda_1'\,(\overline{e}_L\,\vev{\phi_{(3)}^0})_{\underline{1}'}\,e_R''
 + \lambda_1''\,(\overline{e}_L\,\vev{\phi_{(3)}^0})_{\underline{1}''}\,e_R' + \text{h.c.}\;;
 \label{eq:SU5_me}
 \\
 m_u:\,\quad
 &
 - \lambda_2\, \overline{u}_L (\vev{\phi_{(3)}^{0*}}\,u_R)_{\underline{1}}
 - \lambda_2'\, \overline{u}_L'' (\vev{\phi_{(3)}^{0*}}\,u_R)_{\underline{1}'}
 - \lambda_2''\,\overline{u}_L' (\vev{\phi_{(3)}^{0*}}\,u_R)_{\underline{1}''}
   + \text{h.c.}\;;\label{eq:SU5_mu}
\\
 \mnud\,:\quad
 &
 -\lambda_2\, (\overline{\nu}_L\,\vev{\phi_{(3)}^{0*}})_{\underline{1}}\,\nu_R
 - \lambda_2'\,(\overline{\nu}_L\,\vev{\phi_{(3)}^{0*}})_{\underline{1}'}\,\nu_R''
 - \lambda_2''\,(\overline{\nu}_L\,\vev{\phi_{(3)}^{0*}})_{\underline{1}''}\,\nu_R' 
 + \text{h.c.}\;.\label{eq:SU5_mnud}
\end{align}
Expanding out the $A_4$ invariants using the results in the appendix, one obtains
%
\begin{equation}
 m_e   = \Uw \widehat{m}_e\;\;;\;\;
 m_u   = - \widehat{m}_u \Uw\;\;;\;\;
 \mnud = -\Uw \widehat{m}_u\;\;;\quad
 \text{ where  }\;
 \Uw = \MatU \;,\label{eq:SU5_me_mu_mnud}
\end{equation}
where $\widehat{m}_{e,u} = 
\text{diag}(\sqrt{3}\lambda_{1,2}\, v_{(3)},\sqrt{3}\lambda_{1,2}'\, v_{(3)},\sqrt{3}\lambda_{1,2}'' \,v_{(3)})$. 
From (\ref{eq:SU5_me_mu_mnud}), we deduce that 
\begin{equation}
 V_{eL}^\dagger = \Uw \;\;,\;\; V_{uL}^\dagger = V_{eR}=I\;\;,\;\;V_{uR} =-\Uw\;,
\end{equation}
and hence
\begin{equation}
 \mnud = -V_{eL}^\dagger\,\widehat{m}_u\;.
\end{equation}
Putting this into (\ref{eq:ML2}) gives 
\begin{align}
\mnuhat 
 &\simeq V_{\nu}\, V_{eL}^\dagger\,\widehat{m}_u \,M_R^{-1} (V_{eL}^\dagger\,\widehat{m}_u)^T V^{T}_{\nu}\;,
\nonumber\\
 &= \Upmns^\dagger \,\widehat{m}_u \,M_R^{-1}\,\widehat{m}_u \,\Upmns^{*}\;,
 \nonumber\\
\intertext{and hence we arrive at}
M_R 
 &\simeq  \widehat{m}_{u}\, \Upmns^*\, \mnuhat^{-1}\, \Upmns^{\dagger}\, \widehat{m}_{u}\;.
\label{eq:SU5_main_result}
\end{align}
Returning to Eq.~(\ref{eq:SU5_Lag2}), if we expand the $Y_{\lambda 3}$- and $Y_{\lambda 4}$-term in flavor space, it becomes apparent that the $d$-quark mass matrix, $m_d$, and the RH Majorana mass matrix, $M_R$, are both arbitrary complex symmetric matrices. Consequently, the diagonalization matrices $V_{dL}^\dagger$ and $V_\nu$ (since $\mnu$ is a function of $M_R$) are both arbitrary unitary matrices in this model. This implies that the model places no restrictions on the neutrino mixing matrix, $\Upmns = V_{eL} V_\nu^\dagger =\Uw V_\nu^\dagger$, and the quark mixing matrix, $\Uckm = V_{uL} V_{dL}^\dagger =V_{dL}^\dagger$, and so one simply sets them to match the experimental values.

\subsection{Relating $\mnud$ to $\widehat{m}_d$ via a quark-lepton symmetric model}

Next, we construct a slightly more complicated model within the framework of a discrete quark-lepton symmetry  \cite{QL_symmodel}. As well as the usual $A_4$ flavor symmetry, we also introduce an additional unbroken $Z_2$ global symmetry to forbid certain interaction terms in the Lagrangian. The symmetry group is
\begin{align}
 G_2
    &= G_{q\ell} \times A_4 \times Z_2\;,\nonumber\\
    &= \underbrace{SU(3)_\ell \otimes SU(3)_q}_{Z_{QL}} \otimes SU(2)_L \otimes U(1)_X 
    \times A_4 \times Z_2
    \;,\\
    &\supset (SU(2)_\ell\otimes U(1)_T) \otimes SU(3)_q \otimes SU(2)_L \otimes U(1)_X 
    \times A_4 \times Z_2\;, 
\end{align}
where $Z_{QL}$ is the discrete quark-lepton symmetry that relates $SU(3)_{\ell} \leftrightarrow SU(3)_q$ while hypercharge $Y$ is given by a linear function of $X$ and $T$. The field contents are
\begin{align}
 &F_L = \twocMat{N_L}{E_L} \sim (3,1,2,-1/3)(\rept)(1) \quad &\overset{Z_{QL}}{\longleftrightarrow} &\quad
 &Q_L = \twocMat{u_L}{d_L} \sim (1,3,2,1/3)(\rept)(1)\;,\nonumber\\
 &E_R \sim (3,1,1,4/3)(\rep\oplus \repp\oplus \reppp)(1) \quad & \longleftrightarrow &\quad
 &u_R \sim (1,3,1,-4/3)(\rep\oplus \repp\oplus \reppp)(1)\;,\nonumber\\
 &N_R \sim (3,1,1,2/3)(\rep\oplus \repp\oplus \reppp)(-1) \quad & \longleftrightarrow &\quad
 &d_R \sim (1,3,1,-2/3)(\rep\oplus \repp\oplus \reppp)(-1)\;,\nonumber\\
 &\chi_1^{(0)} \sim (3,1,1,2/3)(\rep\oplus \repp\oplus \reppp)(1) \quad & \longleftrightarrow &\quad
 &\chi_2^{(0)} \sim (1,3,1,-2/3)(\rep\oplus \repp\oplus \reppp)(1)\;,\nonumber\\
 &\chi_1^{(1)} \sim (3,1,1,2/3)(\rep\oplus \repp\oplus \reppp)(-1) \quad & \longleftrightarrow &\quad
 &\chi_2^{(1)} \sim (1,3,1,-2/3)(\rep\oplus \repp\oplus \reppp)(-1)\;,\nonumber\\
 &\phi_1 = \twocMat{\phi_1^0}{\phi_1^-} \sim (1,1,2,-1)(\rept)(1) \quad &\longleftrightarrow &\quad
 &\phi_2 = \twocMat{\phi_2^+}{\phi_2^0} \sim (1,1,2,1)(\rept)(1)\;,\nonumber\\
 &\phi_2^c = \twocMat{\phi_2^{0*}}{-\phi_2^-} \sim (1,1,2,-1)(\rept)(1)
   \quad &\longleftrightarrow &\quad
 &\phi_1^c = \twocMat{\phi_1^+}{-\phi_1^{0*}} \sim (1,1,2,1)(\rept)(1)\;,\nonumber\\
 &\phi_d^c = \twocMat{\phi_d^{0*}}{-\phi_d^-} \sim (1,1,2,-1)(\rept)(-1)
   \quad &\longleftrightarrow &\quad
  &\phi_d = \twocMat{\phi_d^+}{\phi_d^0} \sim (1,1,2,1)(\rept)(-1)\;,\nonumber\\
 &\Delta_1 \sim (\overline{6}_s,1,1,-4/3)(\rep\oplus \repp\oplus \reppp)(1) \quad &\longleftrightarrow &\quad
 &\Delta_2 \sim (1,\overline{6}_s,1,4/3)(\rep\oplus \repp\oplus \reppp)(1)\;,
\end{align}
where 
\begin{equation}
E_{L,R} = \threecMat{E_{1L,R}}{E_{2L,R}}{e_{L,R}}\;\;,\;\;
N_{L,R} = \threecMat{N_{1L,R}}{N_{2L,R}}{\nu_{L,R}}\;\;\text{are triplets in $SU(3)_\ell$ space}\;.
\end{equation}
$E_{1L,R}, E_{2L,R}, N_{1L,R}, N_{2L,R}$ are exotic leptonic-color partners of the usual leptons. 
The discrete $Z_{QL}$ symmetry is broken and these exotic leptons gain mass when $\chi_1^{(0,1)}$ picks up a nonzero VEV:
\begin{equation}
 \vev{\chi_1^{(0,1)}} = \threecMat{0}{0}{v_\chi^{(0,1)}}
 \qquad\text{while }\quad \vev{\chi_2^{(0,1)}} = 0\;.
\end{equation}
We arrange $\vev{\Delta_1} \neq 0$ to give a large Majorana mass while keeping $\vev{\Delta_2}=0$. The $\phi$'s will break electroweak symmetry as usual. In order to avoid domain walls\footnote{Cosmological domain walls will form when the discrete quark-lepton symmetry is spontaneously broken. Arranging for this breaking scale to be large allows these observationally unacceptable topological defects to be inflated away \cite{Lew:1992rr}.} and allow the implementation of the seesaw mechanism, we demand the following hierarchy for the energy scales:
\begin{equation}
 \vev{\chi_1^{(0,1)}} > T_\text{inflation} > \vev{\Delta_1} \gg \vev{\phi_1}
 \simeq \vev{\phi_2} \simeq \vev{\phi_d} = \order{10^2} \text{ GeV}\;.
 \label{eq:energy_scale}
\end{equation}
Overall, the $G_2$ invariant interaction Lagrangian takes the form:
\begin{align}
 -\mathcal{L} &=
  \left[\lambda_{f1} \left(\overline{F^c_L}_{\alpha}\, F_{L\beta}\, \chi_{1\gamma}^{(0)}
  +\overline{Q^c_L}_{\alpha}\, Q_{L\beta}\, \chi_{2\gamma}^{(0)}
  \right)
  +\lambda_{f2} \left(\overline{E^c_R}_{\alpha}\, N_{R\beta}\, \chi_{1\gamma}^{(1)}
  +\overline{u^c_R}_{\alpha}\, d_{R\beta}\, \chi_{2\gamma}^{(1)}
  \right)\right]\epsilon^{\alpha\beta\gamma}
  \nonumber\\
  &\qquad
  + \lambda_{g1}\left(\overline{Q}_L u_R \phi_1 +\overline{F}_L E_R \phi_2\right)
  + \lambda_{g2}\left(\overline{Q}_L u_R \phi_2^c +\overline{F}_L E_R \phi_1^c\right)
  + \lambda_{g3}\left(\overline{Q}_L d_R \phi_d +\overline{F}_L N_R \phi_d^c\right)\nonumber\\
  &\qquad
  + \lambda_{h1}\left(\overline{N^c_R}_{\alpha}\, N_{R\beta}\, \Delta_1^{\alpha\beta}
    +\overline{d^c_R}_{\alpha}\, d_{R\beta}\, \Delta_2^{\alpha\beta}\right)
    +\text{h.c.}\;,  \label{QL:Lag}
\end{align}
where $\alpha, \beta, \gamma$ are $SU(3)_{\ell \text{ or } q}$ indices and the terms proportional to $\lambda_{f1,2}$ are the mass terms for the exotic fermions. From (\ref{QL:Lag}) and taking $\vev{\phi_1^0} = v_1, \vev{\phi_2^0} = v_2$ and $\vev{\phi_d^0}\equiv \vev{\phi_d^{0*}} = v_d$, we expect the following mass relations:
\begin{align}
 m_u &= \lambda_{g1} v_1 + \lambda_{g2} v_2^*\;, \; & m_d &= \lambda_{g3} v_d\;,\\
 m_e &= \lambda_{g1} v_2 - \lambda_{g2} v_1^*\;, \;  & \mnud &= \lambda_{g3} v_d \;.
\end{align}
So, in general, $m_e\neq m_u$ but $\mnud=m_d$. Writing out the $A_4$ structure for the above matrices, we have:
\begin{align}
m_e\;:\quad
 &g_1\, (\overline{e}_L\,\vev{\phi_{2}^0})_{\underline{1}}\,e_R
 + g_1'\,(\overline{e}_L\,\vev{\phi_{2}^0})_{\underline{1}'}\,e_R''
 + g_1''\,(\overline{e}_L\,\vev{\phi_{2}^0})_{\underline{1}''}\,e_R' \nonumber\\
&\quad
 -g_2\, (\overline{e}_L\,\vev{\phi_{1}^{0*}})_{\underline{1}}\,e_R
 - g_2'\,(\overline{e}_L\,\vev{\phi_{1}^{0*}})_{\underline{1}'}\,e_R''
 - g_2''\,(\overline{e}_L\,\vev{\phi_{1}^{0*}})_{\underline{1}''}\,e_R'
 + \text{h.c.}\;,\label{eq:QL_me}\\
m_u\;:\quad 
&g_1\, (\overline{u}_L\,\vev{\phi_{1}^0})_{\underline{1}}\,u_R
 + g_1'\,(\overline{u}_L\,\vev{\phi_{1}^0})_{\underline{1}'}\,u_R''
 + g_1''\,(\overline{u}_L\,\vev{\phi_{1}^0})_{\underline{1}''}\,u_R' \nonumber\\
&\quad
 +g_2\, (\overline{u}_L\,\vev{\phi_{2}^{0*}})_{\underline{1}}\,u_R
 + g_2'\,(\overline{u}_L\,\vev{\phi_{2}^{0*}})_{\underline{1}'}\,u_R''
 + g_2''\,(\overline{u}_L\,\vev{\phi_{2}^{0*}})_{\underline{1}''}\,u_R'
+ \text{h.c.}\;,\label{eq:QL_mu}\\
m_d\;:\quad 
  &g_3\, (\overline{d}_L\,\vev{\phi_{d}^0})_{\underline{1}}\,d_R
 + g_3'\,(\overline{d}_L\,\vev{\phi_{d}^0})_{\underline{1}'}\,d_R''
 + g_3''\,(\overline{d}_L\,\vev{\phi_{d}^0})_{\underline{1}''}\,d_R'+ \text{h.c.}\;,
 \label{eq:QL_md}\\
\mnud\;:\quad 
 &g_3\, (\overline{\nu}_L\,\vev{\phi_{d}^{0*}})_{\underline{1}}\,\nu_R
 + g_3'\,(\overline{\nu}_L\,\vev{\phi_{d}^{0*}})_{\underline{1}'}\,\nu_R''
 + g_3''\,(\overline{\nu}_L\,\vev{\phi_{d}^{0*}})_{\underline{1}''}\,\nu_R'+ \text{h.c.}\;.
\end{align}
Choosing the vacuum patterns: $\vev{\phi_{1,2}^{0(*)}} = (v_{1,2}^{(*)},v_{1,2}^{(*)},v_{1,2}^{(*)})\,, \vev{\phi_d^0}\equiv \vev{\phi_d^{0*}} =(v_d,v_d,v_d)$ and following the $A_4$ rules in the appendix, we get
%
\begin{align}
&m_e   = \Uw \widehat{m}_e\;\;,\;\;
 m_u   = \Uw \widehat{m}_u \;\;,\;\;
 m_d   = \mnud = \Uw \widehat{m}_d  \;\;,
 \label{eq:QL_me_mu_md_mnud}\\
\text{i.e. }&\;
 V_{eL}^\dagger = V_{uL}^\dagger = V_{dL}^\dagger = \Uw\;,\; 
 V_{eR}=V_{uR}=V_{dR}=I\;,\label{eq:QL_Vs}
\end{align}
where 
$\widehat{m}_{e} = \text{diag}
(\sqrt{3}(g_1 v_2 -g_2 v_1^*),\sqrt{3}(g_1' v_2 -g_2' v_1^*),\sqrt{3}(g_1'' v_2 -g_2'' v_1^*))$, 
$\widehat{m}_{u} = \text{diag}
(\sqrt{3}(g_1 v_1 +g_2 v_2^*),\sqrt{3}(g_1' v_1 +g_2' v_2^*),\sqrt{3}(g_1'' v_1 +g_2'' v_2^*))$ and 
$\widehat{m}_{d} = \text{diag}(\sqrt{3}\, g_3\, v_d,\sqrt{3}\, g_3'\, v_d,\sqrt{3}\, g_3''\, v_d)$.
In addition, it can be shown that when the $A_4$ singlets \vev{{\Delta_1^{0}}}, \vev{{\Delta_1^{0}}'} and \vev{{\Delta_1^{0}}''} acquire nonzero VEVs, the resulting neutrino Majorana mass matrix, $M_R$, is an arbitrary complex symmetric matrix. Using the results (\ref{eq:QL_me_mu_md_mnud}) and (\ref{eq:QL_Vs}), we can conclude that in this model
\begin{equation}
M_R 
 \simeq  \widehat{m}_{d}\, \Upmns^*\, \mnuhat^{-1}\, \Upmns^{\dagger}\, \widehat{m}_{d}\;,
\end{equation}
where  $\Upmns = V_{eL} V_\nu^\dagger = \Uw^\dagger V_\nu^\dagger$ which is arbitrary, whilst we have $U_\text{CKM} = V_{uL}V_{dL}^\dagger = \Uw^\dagger\Uw =I$. So, at tree-level, this model predicts no quark mixing . However, since the symmetry enforcing this result is now broken, radiative corrections will generate nonzero quark mixing.  We have not attempted to prove that realistic mixing angles can be obtained, since
our focus in this paper is on the lepton sector. 
It is interesting that the form of the mixing matrices predicted by this model is consistent with small quark mixing ($\Uckm \simeq I$), whereas neutrino mixing ($\Upmns = \Uw^\dagger V_\nu^\dagger$) is large \cite{He:2006dk}. This is because $\Uw^\dagger$ is a trimaximal mixing matrix, and so, unless $V_\nu^\dagger \approx \Uw$, one expects the product of the two would be very dissimilar to the identity.

\subsection{Relating $\mnud$ to $\widehat{m}_e$ via a Left-Right model}

Finally, we consider a Left-Right model \cite{LR_Mohapatra.Pati} with $A_4$ flavor symmetry. The symmetry group is
\begin{equation}
 G_3
    = SU(3)_c \otimes SU(2)_L \otimes SU(2)_R \otimes U(1)_{B-L}
    \times A_4
    \;.
    \label{eq:LR_group}
\end{equation}
Here, the imposition of the discrete $L\leftrightarrow R$ parity symmetry is not necessary, and hence will be omitted for simplicity. The complete list of relevant particle contents for this setup is:
\begin{align}
 &\ell_L =\twocMat{\nu_L}{e_L} \sim (1,2,1,-1)(\rept)\;;\quad\quad\quad\quad\quad\quad
 \ell_R =\twocMat{\nu_R}{e_R} \sim (1,1,2,-1)(\rep\oplus \repp\oplus \reppp)\;; \nonumber\\
&\Phi_\ell 
    = \begin{pmatrix} \phi^{0}& \phi^{+}\\ \phi^{-}& -\phi^{0*}\end{pmatrix}
  \sim (1,2,\overline{2},0)(\rept)\;;\quad\quad\quad\quad
\widetilde{\Phi}_\ell 
    = \tau_2\Phi_\ell^* \tau_2 =
  \begin{pmatrix} -\phi^{0}& -\phi^{+}\\ -\phi^{-}& \phi^{0*}\end{pmatrix}
  \sim (1,2,\overline{2},0)(\rept)\;; \nonumber\\
 &q_L =\twocMat{u_L}{d_L} \sim (3,2,1,1/3)(\rept)\;;\quad\quad\quad\quad\quad\quad
 q_R =\twocMat{u_R}{d_R} \sim (3,1,2,1/3)(\rept)\;;\nonumber\\
 &\Phi_q = \begin{pmatrix} \phi^{0}_{A}& \phi^{+}_{B}\\
                              \phi^{-}_{A}& \phi^{0}_{B}\end{pmatrix}
  \sim (1,2,\overline{2},0)(\rep\oplus \repp\oplus \reppp)\;;\quad
 \widetilde{\Phi}_q 
    = \tau_2\Phi_q^* \tau_2 =
  \begin{pmatrix} \phi^{0*}_{B}& -\phi^{+}_{A}\\
                  -\phi^{-}_{B}& \phi^{0*}_{A}\end{pmatrix}
  \sim (1,2,\overline{2},0)(\rep\oplus \repp\oplus \reppp)\;; \nonumber\\
 &\Delta_R = \begin{pmatrix} \delta^{+}/\sqrt{2}& \delta^{++}\\
                              \delta^{0}& -\delta^{+}/\sqrt{2}\end{pmatrix}
  \sim (1,1,3,2)(\rep\oplus \repp\oplus \reppp)\;,
\end{align}
where we have deliberately embedded the same Higgs doublet into $\Phi_\ell$ to form a \emph{real} bidoublet. In matrix form, the $G_3$ invariant Lagrangian has the following terms:
\begin{align}
 -\mathcal{L} &=
      \lambda_{y1}\,\overline{\ell}_L\,\Phi_\ell\,\ell_R
   + \widetilde{\lambda}_{y1}\,\overline{\ell}_L\,\widetilde{\Phi}_\ell\,\ell_R
   + \lambda_{y2}\,\overline{q}_L\,\Phi_\ell\,q_R
   + \widetilde{\lambda}_{y2}\,\overline{q}_L\,\widetilde{\Phi}_\ell\,q_R
   + \lambda_{y3}\,\overline{q}_L\,\Phi_q\,q_R
   + \widetilde{\lambda}_{y3}\,\overline{q}_L\,\widetilde{\Phi}_q\,q_R
     \nonumber\\
   &\quad
   + \lambda_{y4}\,\overline{\ell_R^c}\, i \tau_2\, \Delta_R\, \ell_R
  +\text{h.c.}\;. \label{LR:Lag}
\end{align}
When the symmetry is broken spontaneously by the nonzero VEVs,
\begin{equation}
 \vev{\Phi_\ell} = \begin{pmatrix} v_\ell & 0\\ 0 & -v_\ell
                   \end{pmatrix}
                 \equiv - \vev{\widetilde{\Phi}_\ell}
                   \;;
  \quad
 \vev{\Phi_q} = \begin{pmatrix} v_A & 0\\ 0 & v_B
                   \end{pmatrix}\;;
  \quad
 \vev{\widetilde{\Phi}_q} = \begin{pmatrix} v_B^* & 0\\ 0 & v_A^*
                   \end{pmatrix}\;;
  \quad
 \vev{\Delta_R} = \begin{pmatrix} 0 & 0\\ v_\delta & 0
                   \end{pmatrix}\;,
\end{equation}
where $v_\ell \in \mathbb{R}$ and $\order{v_\delta} \gg \order{v_{\ell,A,B}}$, we obtain mass relations of the form:
\begin{align}
 m_u &= (\lambda_{y2} -\widetilde{\lambda}_{y2}) \,v_\ell +
       \lambda_{y3} \,v_A +\widetilde{\lambda}_{y3} \,v_B^* \;,
 &\mnud = (\lambda_{y1} -\widetilde{\lambda}_{y1}) \,v_\ell\;,\\
 m_d &= -(\lambda_{y2} -\widetilde{\lambda}_{y2}) \,v_\ell +
       \lambda_{y3} \,v_B +\widetilde{\lambda}_{y3} \,v_A^* \;,
 &m_e = -(\lambda_{y1} -\widetilde{\lambda}_{y1}) \,v_\ell\;,
\end{align}
In flavor space, the charged-lepton and neutrino Dirac-mass terms become
\begin{align}
 m_e\;:\qquad
 &-\left[
    y_1 \,(\overline{e}_L\,\vev{\phi^{0*}})_{\underline{1}}\,e_R
  + y_1' \,(\overline{e}_L\,\vev{\phi^{0*}})_{\underline{1}'}\,e_R''
  + y_1'' \,(\overline{e}_L\,\vev{\phi^{0*}})_{\underline{1}''}\,e_R'
 \right]\nonumber\\
 &\qquad
  + \widetilde{y}_1 \,(\overline{e}_L\,\vev{\phi^{0*}})_{\underline{1}}\,e_R
  + \widetilde{y}_1' \,(\overline{e}_L\,\vev{\phi^{0*}})_{\underline{1}'}\,e_R''
  + \widetilde{y}_1'' \,(\overline{e}_L\,\vev{\phi^{0*}})_{\underline{1}''}\,e_R'
  +\text{h.c.}\;,\label{eq:LR_me}\\
 \mnud\;:\qquad
 &
    y_1 \,(\overline{\nu}_L\,\vev{\phi^{0}})_{\underline{1}}\,\nu_R
  + y_1' \,(\overline{\nu}_L\,\vev{\phi^{0}})_{\underline{1}'}\,\nu_R''
  + y_1'' \,(\overline{\nu}_L\,\vev{\phi^{0}})_{\underline{1}''}\,\nu_R'
  \nonumber\\
&\qquad
  -\left[ \widetilde{y}_1 \,(\overline{\nu}_L\,\vev{\phi^{0}})_{\underline{1}}\,\nu_R
  + \widetilde{y}_1' \,(\overline{\nu}_L\,\vev{\phi^{0}})_{\underline{1}'}\,\nu_R''
  + \widetilde{y}_1'' \,(\overline{\nu}_L\,\vev{\phi^{0}})_{\underline{1}''}\,\nu_R'
  \right]
  +\text{h.c.}\;.\label{eq:LR_mnud}
\end{align}  
%
Taking $\vev{\phi^{0*}} \equiv \vev{\phi^0}
  = (v_\ell, v_\ell, v_\ell)$ and then comparing Eqs.~(\ref{eq:LR_me}) and (\ref{eq:LR_mnud}), one gets
\begin{equation}
 m_e = \Uw \widehat{m}_e\;,\quad \mnud = -\Uw \widehat{m}_e = -V_{eL}^\dagger \widehat{m}_e\;,\;
 \label{eq:LR_relation}
\end{equation}
where $\widehat{m}_e = \text{diag}(\sqrt{3}(-y_1 +\widetilde{y}_1)v_\ell\;, \sqrt{3}(-y_1' +\widetilde{y}_1')v_\ell\;, \sqrt{3}(-y_1''+\widetilde{y}_1'')v_\ell)$. Whereas the neutrino Majorana mass matrix is a general complex symmetric just like in our other examples, the quark mass matrices have a special form. For $m_u$, the expanded Lagrangian,
\begin{align}
  &
   y_{2s}\,(\overline{u}_L\,u_R)_{\rept s} \,\vev{\phi^{0}}
  +y_{2a}\,(\overline{u}_L\,u_R)_{\rept a} \,\vev{\phi^{0}}
  -\widetilde{y}_{2s}\,(\overline{u}_L\,u_R)_{\rept s} \,\vev{\phi^{0}}
  -\widetilde{y}_{2a}\,(\overline{u}_L\,u_R)_{\rept a} \,\vev{\phi^{0}} 
  +y_3\,(\overline{u}_L\,u_R)_{\rep} \,\vev{\phi^{0}_A}
\nonumber\\
  &\quad  
  +y_3'\,(\overline{u}_L\,u_R)_{\repp} \,\vev{{\phi^{0}_A}''}
  +y_3''\,(\overline{u}_L\,u_R)_{\reppp} \,\vev{{\phi^{0}_A}'} 
  +\widetilde{y}_3\,(\overline{u}_L\,u_R)_{\rep} \,\vev{\phi^{0*}_B}
  +\widetilde{y}_3'\,(\overline{u}_L\,u_R)_{\repp} \,\vev{{\phi^{0*}_B}''}
  +\widetilde{y}_3''\,(\overline{u}_L\,u_R)_{\reppp} \,\vev{{\phi^{0*}_B}'}+\text{h.c.}\;,
\end{align}
gives rise to a mass matrix of the form
\begin{equation}
 m_u=\3by3Mat{A_{1}}{B_{+}}{B_{-}}
             {B_{-}}{A_{2}}{B_{+}}
             {B_{+}}{B_{-}}{A_{3}} \;, \label{eq:LR_mu}
\end{equation}
while it can be shown that mass matrix $m_d$ also has a similar structure: 
\begin{equation}
 m_d=\3by3Mat{C_{1}}{-B_{+}}{-B_{-}}
             {-B_{-}}{C_{2}}{-B_{+}}
             {-B_{+}}{-B_{-}}{C_{3}} \;,\label{eq:LR_md} 
\end{equation}
where $A_{1,2,3}$ ,$C_{1,2,3}$ and $B_{+,-}$ are complicated functions of the VEVs and Yukawa couplings. 
Equations (\ref{eq:LR_mu}) and (\ref{eq:LR_md}) imply that the diagonalization matrices $V_{uL}$ and $V_{dL}$ are not completely arbitrary. However, it is easy to see that there are enough degrees of freedom in the resulting $\Uckm = V_{uL}V_{dL}^\dagger$ such that experimental data can be fitted. Returning to (\ref{eq:LR_relation}), it is clear that the main prediction of this model is 
\begin{equation}
M_R 
 \simeq  \widehat{m}_{e}\, \Upmns^*\, \mnuhat^{-1}\, \Upmns^{\dagger}\, \widehat{m}_{e}\;,
\end{equation}
where  $\Upmns = V_{eL} V_\nu^\dagger = \Uw^\dagger V_\nu^\dagger$ is \emph{a priori} arbitrary and to be fitted to the experimental observations.

\section{Phenomenology}
\label{Sec:phenomenology}

The general conclusion from the previous section is that it is possible to use symmetries to construct the relation
\begin{equation}
 M_R \simeq	
  \widehat{m}_f\,\Upmns^*\, \mnuhat^{-1}\, \Upmns^{\dagger}\, \widehat{m}_{f}\;,\quad 
  f= e, d \text{ or } u\;,
\label{eq:ph_MR}
\end{equation}
that links the high-energy seesaw sector to low-energy observables. Using the current experimental data on quarks and leptons, the properties of the heavy RH Majorana neutrinos in these models can therefore be inferred directly, and interesting consequences may arise. 

Whilst the mixing matrix $\Upmns$ can be in general written as
\begin{equation}\label{eq:UPMNS_param}
 \Upmns = 
 \3by3Mat{c_{12}c_{13}}{s_{12}c_{13}}{s_{13}\,e^{-i\delta}}
 {-s_{12}c_{23}-c_{12}s_{23}s_{13}\,e^{i\delta}}
 {c_{12}c_{23}-s_{12}s_{23}s_{13}\,e^{i\delta}}
 {s_{23}c_{13}}
 {-s_{12}s_{23}+c_{12}c_{23}s_{13}\,e^{i\delta}}
 {c_{12}s_{23}+s_{12}c_{23}s_{13}\,e^{i\delta}}
 {-c_{23}c_{13}} \Matdiag{e^{i\alpha_1/2}}{e^{i\alpha_2/2}}{1},
\end{equation}
where $s_{mn}=\sin\theta_{mn}, c_{mn}=\cos\theta_{mn}$, $\delta$ is the $CP$-violating Dirac phase, and $\alpha_1$ and $\alpha_2$ denote the two Majorana phases, it is often more convenient to absorb the Majorana phases into $\mnuhat$ in (\ref{eq:ph_MR}) and allow the $m_i$'s to be complex masses instead. When numerical analysis is required, we use the best fit values with $1\sigma$ errors for the mixing angles \cite{Schwetz:2008er}:
\begin{equation} \label{eq:bestfit_mixing}
 \sin^2\theta_{12} = 0.304^{+0.022}_{-0.016}\;,\quad
 \sin^2\theta_{23} = 0.50^{+0.07}_{-0.06}\;,\quad
 \sin^2\theta_{13} = 0.01^{+0.016}_{-0.011}\;.
\end{equation}
But for our analytical work, we assume that $\Upmns$ has an exact tribimaximal form \cite{TBmixing}, with  
\begin{equation}\label{eq:TB_values}
 \sin^2\theta_{12} = \frac{1}{3}\;,\quad
 \sin^2\theta_{23} = \frac{1}{2}\;,\quad
 \sin^2\theta_{13} = 0\;.
\end{equation}
The inputs to the light neutrino mass matrix $\mnuhat$ are restricted by the squared-mass differences:
\begin{equation}\label{eq:mass_sq_diff}
 \Delta\msol^2 = 7.65^{+0.23}_{-0.20} \times 10^{-5}\;\text{eV}^2 \;,\quad 
 \Delta\matm^2 = 2.40^{+0.12}_{-0.11} \times 10^{-3}\;\text{eV}^2 \;,
\end{equation}
obtained from neutrino oscillation experiments \cite{neutrinos_exp, mixingdata, Schwetz:2008er} and the cosmological bound on the sum of all neutrino masses: $\sum_i |m_i| \lesssim 0.61$~eV (95\% C.L.)  \cite{WMAP} which implies an absolute upper limit of
\begin{equation}\label{eq:nu_mass_limit}
 |m_i| < 0.2 \text{ eV} \;\;(95\% \text{ C.L.}) \quad \text{for all }\; i\;. 
\end{equation}
In the following, we study (\ref{eq:ph_MR}) by taking a generic form for $\widehat{m}_f \equiv \text{diag}(\xi_1,\xi_2,\xi_3)$ where $\xi_1 \ll \xi_2 \ll \xi_3$ is assumed. 
It is obvious that once $\widehat{m}_f$ has been chosen (i.e. $\xi_i$'s are known), only $\delta,\alpha_1,\alpha_2$ and $|m_1|$ (or $|m_3|$ for the inverted hierarchy case) can 
potentially change the form of $M_R$ and its eigenvalue spectrum. Moreover, if $\theta_{13} \simeq 0$, it is expected that the Dirac phase, $\delta$, would not play a significant role.\footnote{It should be pointed out that when 13-mixing is nonzero, say at the best fit value of $5.7^\circ$, the choice of Dirac phase can influence the mass eigenvalues by almost two orders of magnitude for certain sets of Majorana phases and $|m_{1,3}|$ values, as our parameter space scans have indicated.}

So, to understand the leading behaviors of the mass spectrum for $M_R$, we approximate $\Upmns$ with the tribimaximal form (see (\ref{eq:TB_values})) and absorb $\alpha_{1,2}$ into $m_{1,2}$ respectively. After expanding out the RHS of (\ref{eq:ph_MR}), we have
\begin{equation}\label{eq:Mexact}
 M_R\equiv M_R^T = 
 \3by3Mat
 {\frac{2\xi_1^2}{3m_1}+\frac{\xi_1^2}{3m_2}}
 {-\frac{\xi_1 \xi_2}{3m_1}+\frac{\xi_1\xi_2}{3m_2}}
 {-\frac{\xi_1 \xi_3}{3m_1}+\frac{\xi_1\xi_3}{3m_2}}
 {\cdots}
 {\frac{\xi_2^2}{6m_1}+\frac{\xi_2^2}{3m_2}+\frac{\xi_2^2}{2m_3}}
 {\frac{\xi_2\xi_3}{6m_1}+\frac{\xi_2\xi_3}{3m_2}-\frac{\xi_2\xi_3}{2m_3}}
 {\cdots}
 {\cdots}
 {\frac{\xi_3^2}{6m_1}+\frac{\xi_3^2}{3m_2}+\frac{\xi_3^2}{2m_3}}\;.
\end{equation}
There are two limiting cases of Eq.~(\ref{eq:Mexact}) which can provide important insights into the dependence of the heavy RH Majorana masses $M_i$ on the mass scale of the lightest LH neutrino.
\subsection{Fully hierarchical light neutrinos}

For the normal hierarchy scheme, we have $|m_1| \rightarrow 0$ with $|m_{2,3}|$ related to $|m_1|$ via (\ref{eq:NH}). Therefore, in this limit, we can write Eq.~(\ref{eq:Mexact}) as
\begin{equation}\label{eq:MR_MR0_dM}
 M_R = M_{R0} + \Delta M_R\;, \qquad \text{where }\;\;
 M_{R0}\equiv \3by3Mat
 {\frac{2\xi_1^2}{3m_1}}
 {-\frac{\xi_1 \xi_2}{3m_1}}
 {-\frac{\xi_1 \xi_3}{3m_1}}
 {\cdots}
 {\frac{\xi_2^2}{6m_1}}
 {\frac{\xi_2\xi_3}{6m_1}}
 {\cdots}
 {\cdots}
 {\frac{\xi_3^2}{6m_1}}\;
\end{equation}
is the dominant part of the matrix as $|m_1| \rightarrow 0$, while $\Delta M_R$ is considered to be a small perturbation. Suppose that the true eigenvalues and eigenvectors for $M_R$ can be expressed as 
$E_{i} \equiv E_{i0} + \Delta E_i$ and $u_{i} \equiv u_{i0} + \Delta u_i$ respectively for all $i$, where $M_{R0}\,u_{i0} = E_{i0}\,u_{i0}$ is assumed. Then, perturbation theory implies that the variation in the eigenvalues is given by
\begin{equation}\label{eq:MR_dE_eqn}
 \Delta E_i = u_{i0}^T \cdot (\Delta M_R) \cdot u_{i0}\;, \quad i=1,2 \text{ and } 3\;,
\end{equation}
where $u_{i0}$'s are chosen to be orthonormal to each other. Solving $M_{R0}\,u_{i0} = E_{i0}\,u_{i0}$ for $E_{i0}$, one immediately gets
\begin{equation}\label{eq:MR_E0}
 E_{10} \;, E_{20} = 0\;,\quad E_{30} = \frac{4\xi_1^2+\xi_2^2+\xi_3^2}{6m_1}
 \simeq \frac{\xi_3^2}{6|m_1|}\;,
\end{equation}
and subsequently
\begin{equation}\label{eq:MR_dE}
 \Delta E_1 \simeq \frac{3\xi_1^2}{m_2}\;,\quad
 \Delta E_2 \simeq \frac{2\xi_2^2}{m_3}\;,\quad
 \Delta E_3 \simeq \frac{2\xi_3^2}{m_2}\;,
\end{equation}
 in the limit of $\xi_3 \gg \xi_{1,2}$ and $|m_3| \gg |m_2|$.  
Combining Eqs.~(\ref{eq:MR_E0}) and (\ref{eq:MR_dE}), the heavy RH neutrino masses are
\begin{equation}\label{eq:NH_MR_mass}
 |M_1| \simeq \frac{3\xi_1^2}{|m_2|}\;,\quad
 |M_2| \simeq \frac{2\xi_2^2}{|m_3|}\;,\quad
 |M_3| \simeq \frac{\xi_3^2}{6|m_1|}\;.
\end{equation}
Hence, we can see that due to the large neutrino mixing, the expected correspondence between $m_i$ and the Dirac masses, $m_i \propto \xi_i^2$, no longer holds and that only the largest RH neutrino mass is a function of $|m_1|$.\footnote{These results are consistent with those in references \cite{Akhmedov03, Branco02}.} Substituting in the running fermion masses $m(\mu)$ at $\mu \simeq 10^9 \text{ GeV}$ \cite{running_mass} as typical values for $\xi_i$'s, we have the following predictions for RH neutrino masses in the normal hierarchy case:
\begin{align}
 m_u\;: \quad
  &|M_1| \simeq 5.6\times 10^5 \text{ GeV}\;,\quad
   |M_2| \simeq 5.5\times 10^{9} \text{ GeV}\;,\quad
    |M_3| \gtrsim 2.0\times 10^{14} \text{ GeV}\;, \label{eq:MR_mass_u_case}\\
  m_d\;: \quad
  &|M_1| \simeq 2.3\times 10^6 \text{ GeV}\;,\quad
   |M_2| \simeq 1.1\times 10^8 \text{ GeV}\;,\quad
    |M_3| \gtrsim 3.8\times 10^{10} \text{ GeV}\;, \label{eq:MR_mass_d_case}\\
  m_e\;: \quad
  &|M_1| \simeq 9.0\times 10^4 \text{ GeV}\;,\quad
   |M_2| \simeq 4.8\times 10^8 \text{ GeV}\;,\quad
    |M_3| \gtrsim 5.7\times 10^{10} \text{ GeV}\;. \label{eq:MR_mass_e_case}        
\end{align}
The plots of $M_{1,2,3}$ as a function of $|m_1|$ for the case $\widehat{m}_f=\widehat{m}_u$ and for many different values of $\delta, \alpha_{1,2}$ are shown in Fig.~\ref{fig:mu_M123}. These numerical results validate the trend predicted by the theoretical analysis. The tallest spikes in the diagrams of Fig.~\ref{fig:mu_M123} are locations where level crossing occurs ($M_{1,2}$ or $M_{2,3}$ are quasi-degenerate) for certain special values of Dirac and Majorana phases, an effect that has been previously studied in \cite{Akhmedov03, Nezri00}. Plots of $M_{1,2}$ for the case $\widehat{m}_f=\widehat{m}_{d,e}$ are shown in Fig.~\ref{fig:md_me_M12}.

For the inverted hierarchy scheme, $|m_3| \ll |m_1| \simeq |m_2|$, and hence, we take
\begin{equation}\label{eq:IH_MR0}
 M_{R0}\equiv 
 \3by3Mat
 {0}
 {0}
 {0}
 {\cdots}
 {\frac{\xi_2^2}{2m_3}}
 {\frac{-\xi_2\xi_3}{2m_3}}
 {\cdots}
 {\cdots}
 {\frac{\xi_3^2}{2m_3}}\;,
\end{equation}
which then leads to the following expressions for the $M_R$ masses:
\begin{equation}\label{eq:IH_MR_mass}
 |M_1| \simeq \frac{\xi_1^2}{|m_2|}\;,\quad
 |M_2| \simeq \frac{2\xi_2^2}{|m_2|}\;,\quad
 |M_3| \simeq \frac{\xi_3^2}{2|m_3|}+\frac{\xi_3^2}{2|m_2|}
       \simeq \frac{\xi_3^2}{2|m_3|}
 \;.
\end{equation}
The resulting numerical values for this case are similar to those shown in Eqs.~(\ref{eq:MR_mass_u_case}) to (\ref{eq:MR_mass_e_case}) although they are in general slightly smaller.

\subsection{Quasi-degenerate light neutrinos}

\begin{figure}[tb]
\centering 
\epsfig{file=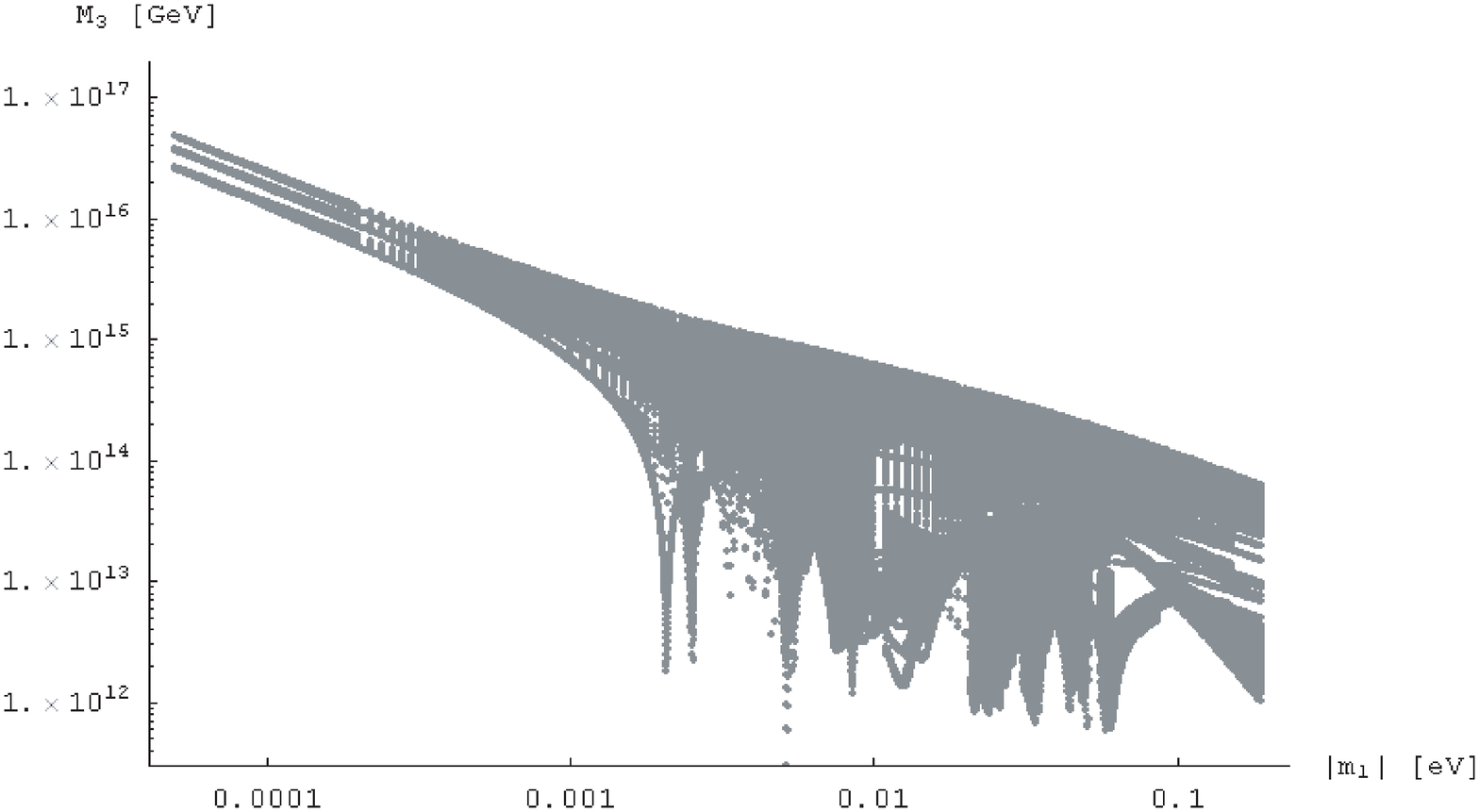, width=12cm}
%
\epsfig{file=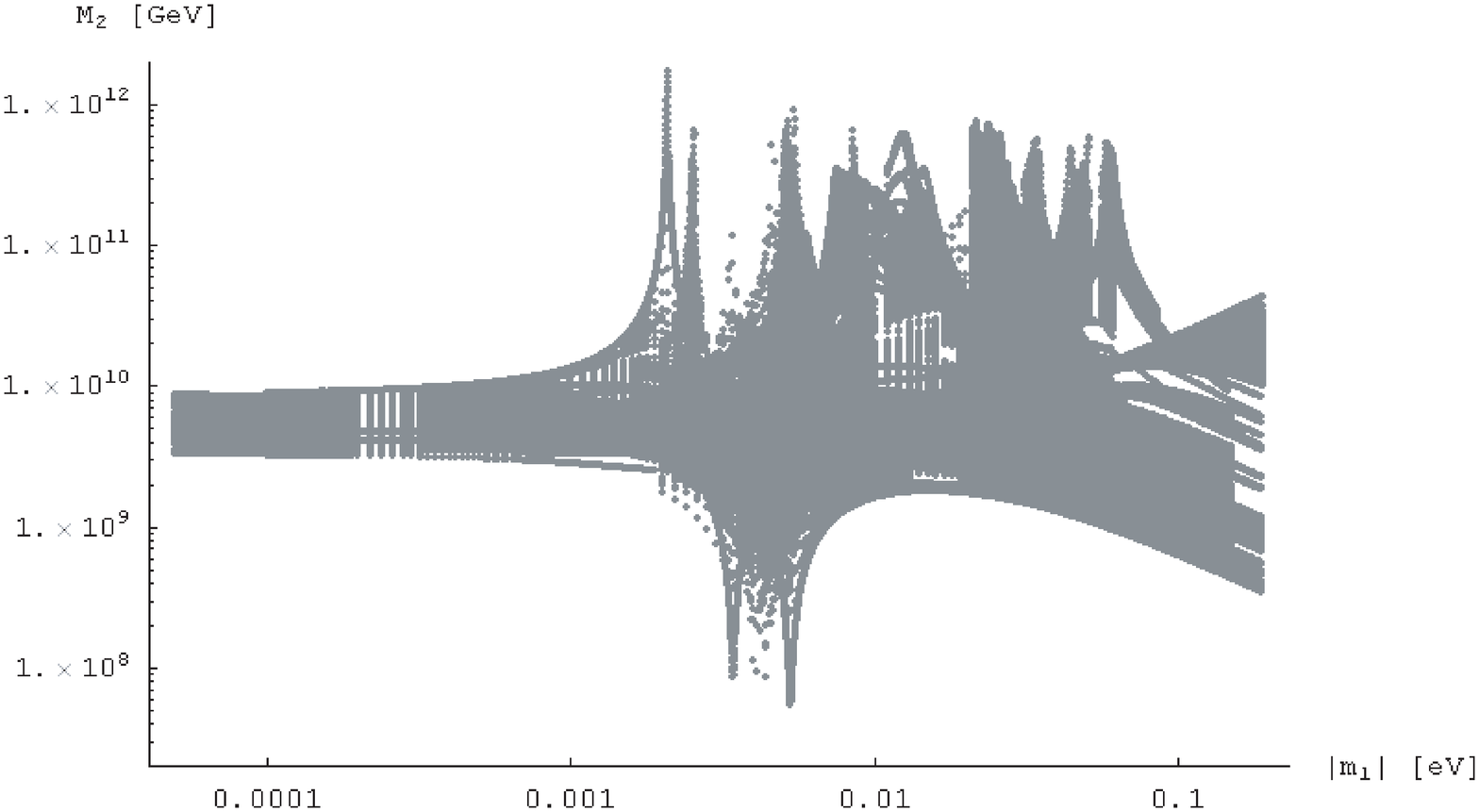, width=12cm}
%
\epsfig{file=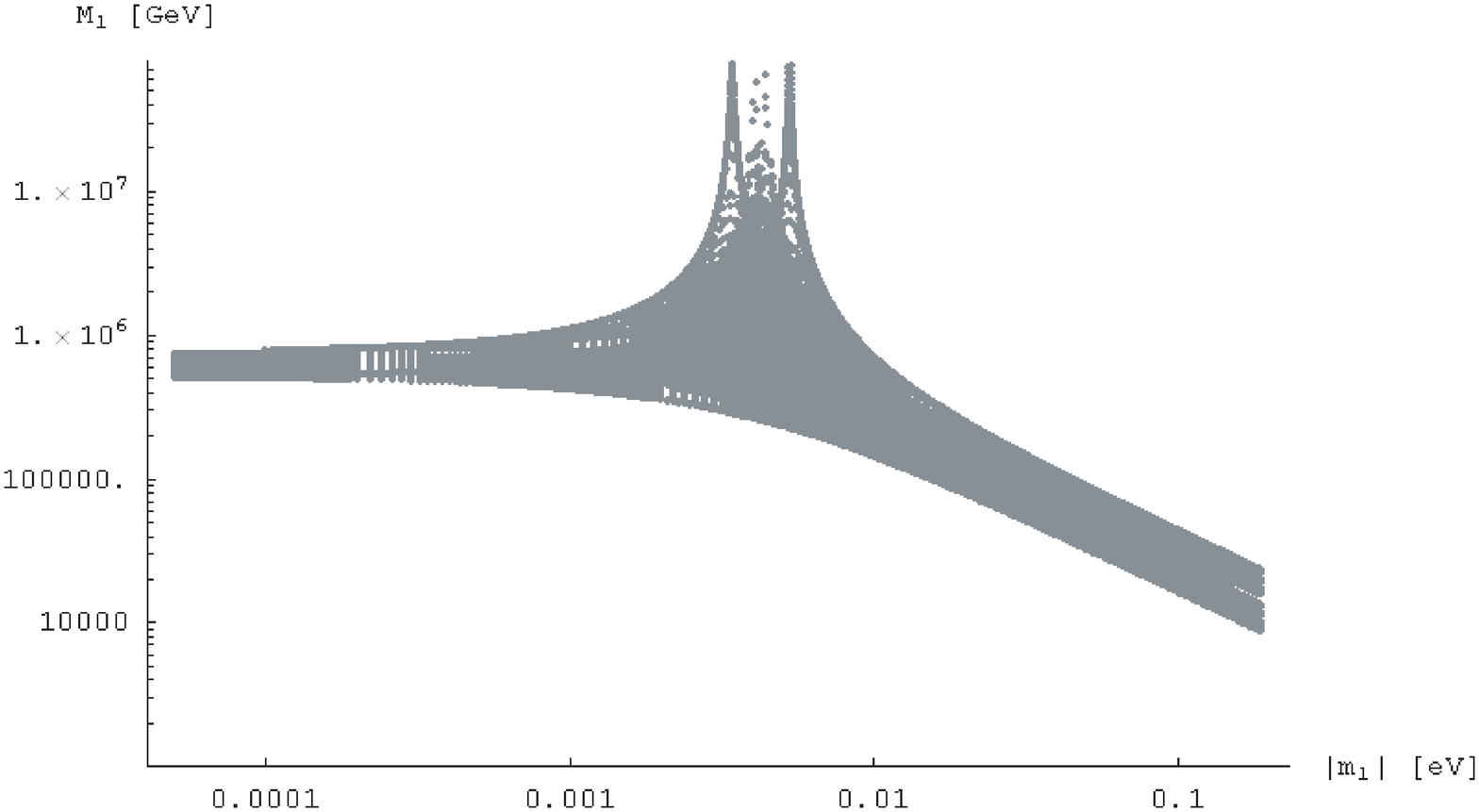, width=12cm}
\caption{Plots of $M_{1,2,3}$ vs. $|m_1|$ in the $\widehat{m}_f=\widehat{m}_u$ case with normal hierarchy for light neutrino masses assumed. 
Input running masses used: $m_u(\mu)=1.3$~MeV, $m_c(\mu)=0.37$~GeV, $m_t(\mu)=1.1\times 10^2$~GeV, where $\mu\simeq 10^9$~GeV.
Each plot contains approximately $3.18\times 10^5$ data points produced by systematically sweeping the $|m_1|$ and $\delta, \alpha_{1,2} \in (0, 2\pi)$ parameter space.}\label{fig:mu_M123}
\end{figure}

\begin{figure}[tb]
\includegraphics[width=0.45\columnwidth]{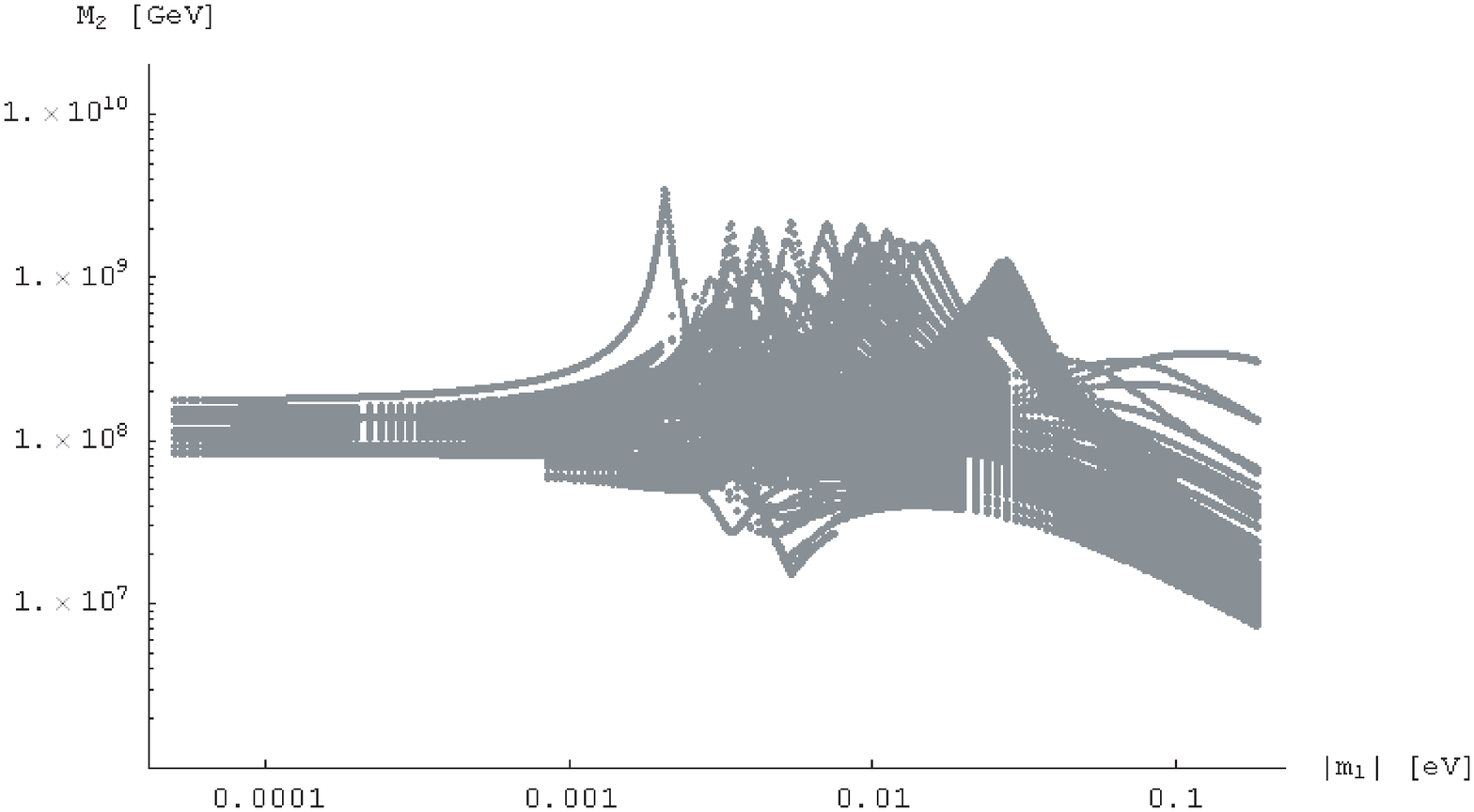}
\includegraphics[width=0.45\columnwidth]{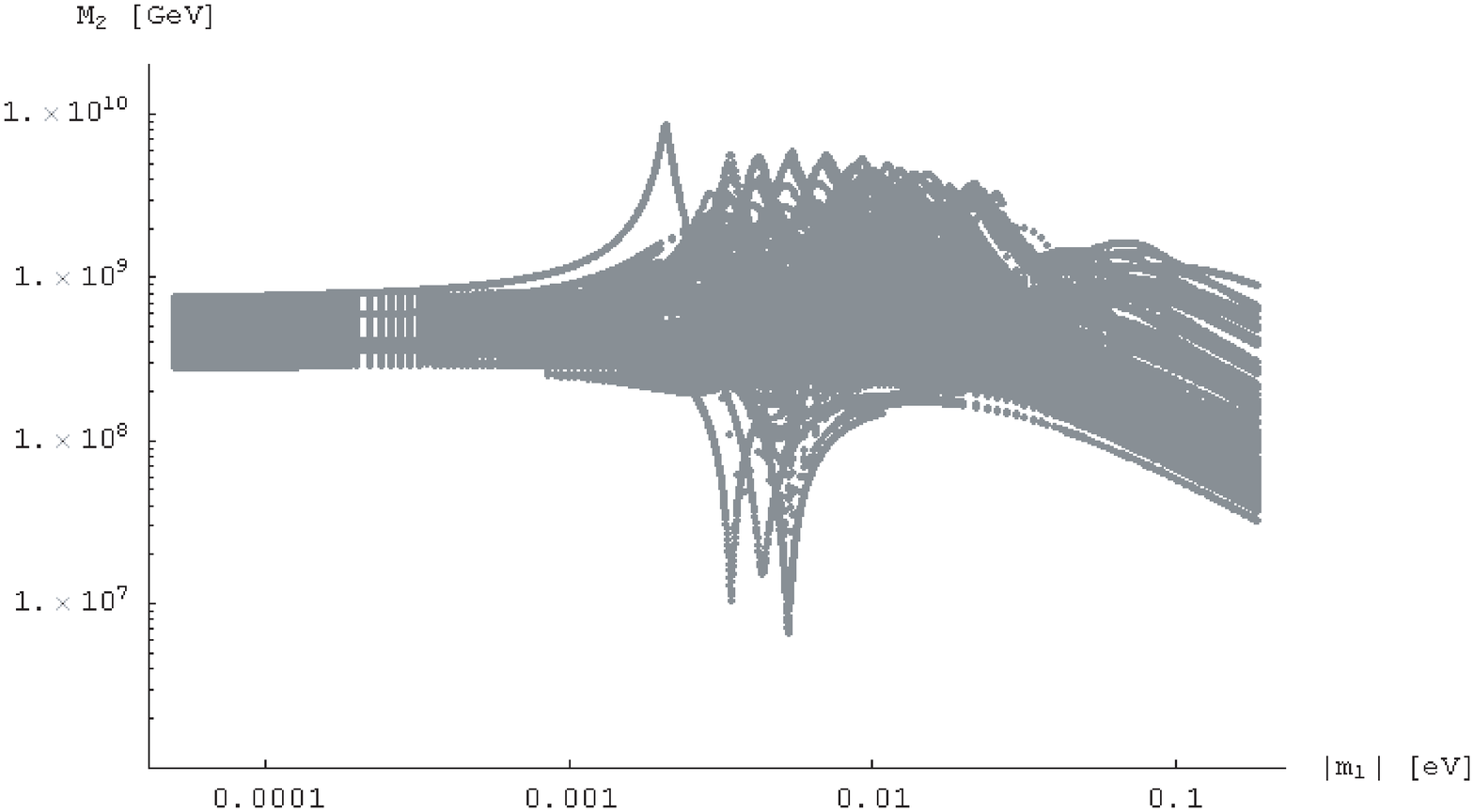}
\includegraphics[width=0.45\columnwidth]{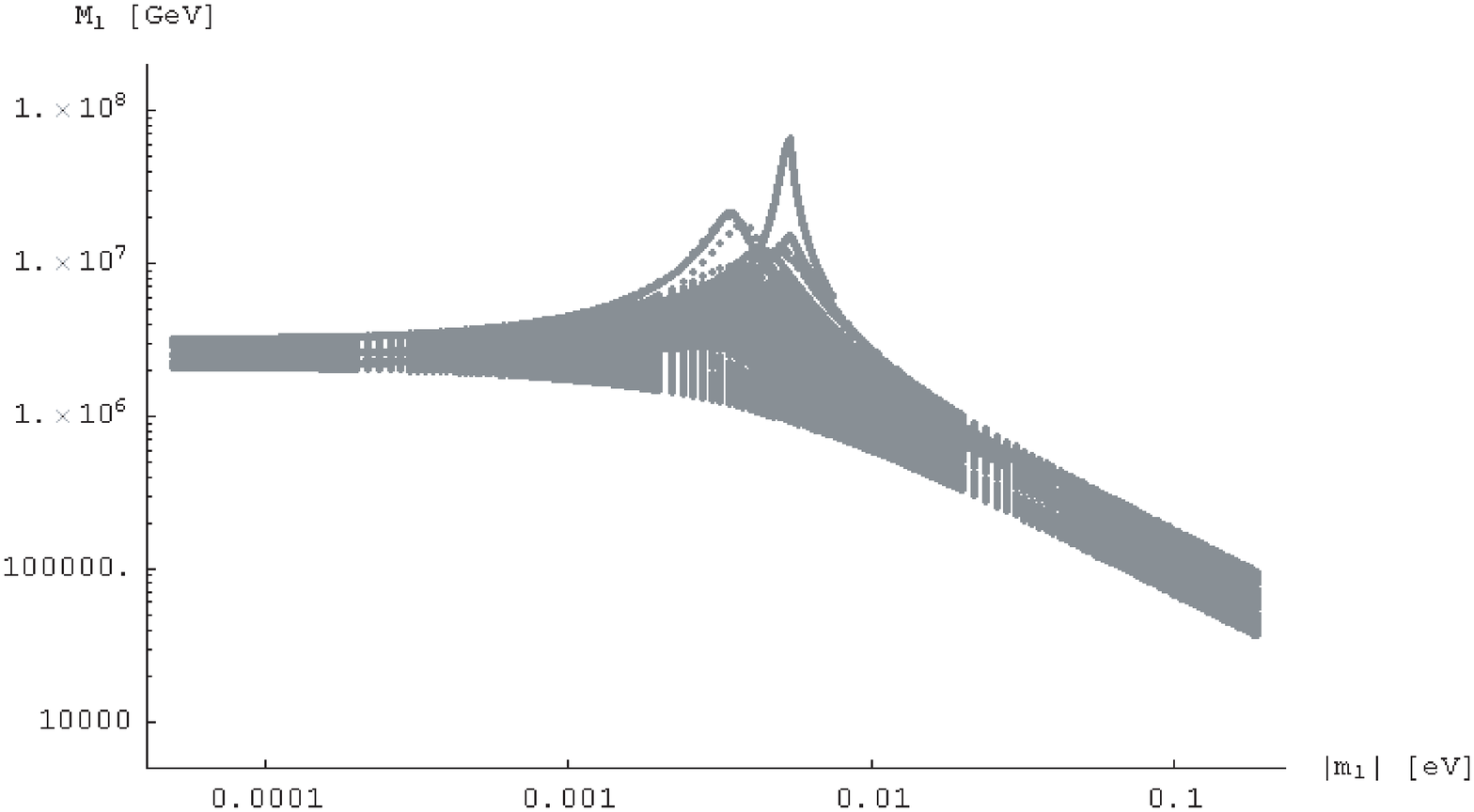}
\includegraphics[width=0.45\columnwidth]{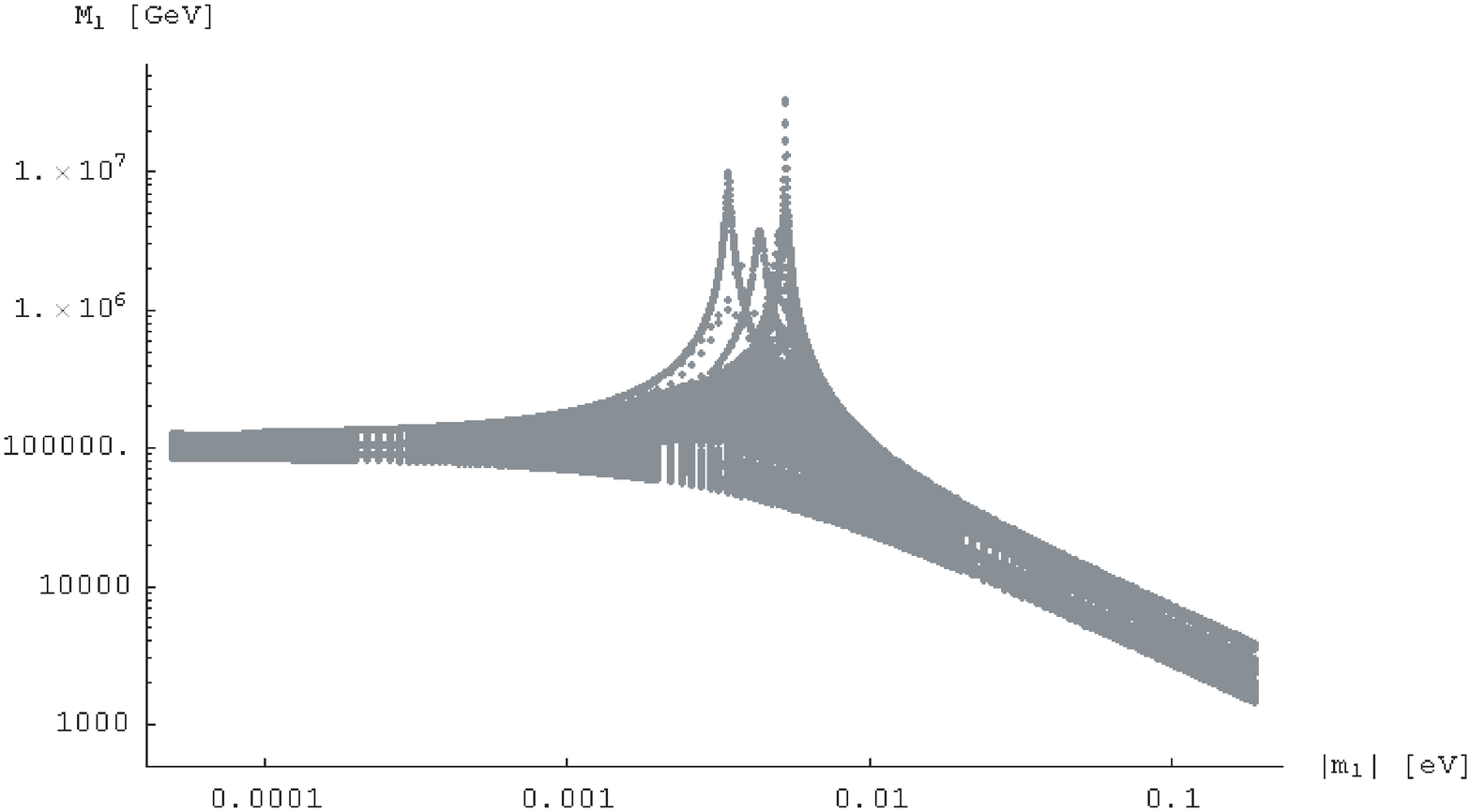}
\caption{Plots of $M_{1,2}$ vs. $|m_1|$ in the $\widehat{m}_f=\widehat{m}_d$ case (LEFT column) and $\widehat{m}_f=\widehat{m}_e$ case (RIGHT column) with normal hierarchy for light neutrino masses assumed. Input running masses used: 
(LEFT) $m_d(\mu)=2.6$~MeV, $m_s(\mu)=52$~MeV, $m_b(\mu)=1.5$~GeV, and 
(RIGHT) $m_e(\mu)=0.52$~MeV, $m_\mu(\mu)=1.1\times 10^2$~MeV, $m_\tau(\mu)=1.8$~GeV, 
where $\mu\simeq 10^9$~GeV.
Each plot contains approximately $1.0\times 10^5$ data points produced by systematically sweeping $|m_1|$ and the $\delta, \alpha_{1,2} \in (0, 2\pi)$ parameter space.}
\label{fig:md_me_M12}
\end{figure}

When the lightest neutrino mass approaches the upper bound of (\ref{eq:nu_mass_limit}), we get $|m_1|\simeq |m_2| \simeq|m_3|$. Assuming that the Majorana phases $\alpha_{1,2}$ are negligible, then Eq.~(\ref{eq:Mexact}) becomes
\begin{equation}\label{eq:QD_Mexact}
 M_R \simeq 
 \3by3Mat
 {\frac{\xi_1^2}{|m_1|}}{0}{0}
 {\cdots}{\frac{\xi_2^2}{|m_1|}}{0}
 {\cdots}{\cdots}{\frac{\xi_3^2}{|m_1|}}\;.
\end{equation}
From this, we can immediately deduce the approximate scale for the $M_i$'s:
\begin{align}
 m_u\;: \quad
  &|M_1| \simeq 8.5\times 10^3 \text{ GeV}\;,\quad
   |M_2| \simeq 6.8\times 10^{8} \text{ GeV}\;,\quad
    |M_3| \simeq 5.9\times 10^{13} \text{ GeV}\;, \label{eq:QDMR_u_case}\\
  m_d\;: \quad
  &|M_1| \simeq 3.4\times 10^4 \text{ GeV}\;,\quad
   |M_2| \simeq 1.3\times 10^7 \text{ GeV}\;,\quad
    |M_3| \simeq 1.1\times 10^{10} \text{ GeV}\;, \label{eq:QDMR_d_case}\\
  m_e\;: \quad
  &|M_1| \simeq 1.4\times 10^3 \text{ GeV}\;,\quad
   |M_2| \simeq 5.9\times 10^7 \text{ GeV}\;,\quad
    |M_3| \simeq 1.7\times 10^{10} \text{ GeV}\;. \label{eq:QDMR_e_case}        
\end{align}
These estimates agree well with the numerical results shown in Fig.~\ref{fig:mu_M123} and \ref{fig:md_me_M12}.

\subsection{Thermal Leptogenesis}

Using the $M_R$ mass spectrum information presented above, several general comments on the possibility of baryon asymmetry generation via thermal leptogenesis for the models discussed in Section~\ref{Sec:models} can be made.
First of all, from the fact that $M_1$ is typically in the range of $10^3 - 10^6$ GeV for all setups, it is clear that conventional leptogenesis where the asymmetry is generated predominantly by the decays of $N_1$'s would not be successful \cite{BDP, Akhmedov03}.  However, there exist other special solutions to the leptogenesis scenario.

As was pointed out earlier, the tall spikes in the plots of Figs.~\ref{fig:mu_M123} and \ref{fig:md_me_M12} indicate that there are regions in the parameter space for these models where $M_{1}$ and $M_2$ become almost degenerate. Consequently, it has been shown in \cite{Akhmedov03} that a sufficient baryon asymmetry can be generated from resonant enhancement \cite{resonant} to the raw $CP$ asymmetry in the decays of $N_1$'s. Furthermore, a similar enhancement to the decay of the next-to-the-lightest RH neutrino $N_2$, when $M_2$ and $M_3$ become degenerate, can also produce the desired asymmetry in principle, as long as washout effects mediated by the lighter $N_1$'s are insufficient \cite{DiBari:2005st}.

Another interesting observation is that, recently, Ref.~\cite{Bari_Riotto} discussed the possibility of successful leptogenesis (without the need for resonant enhancement) in models with $SO(10)$-inspired mass relations which have properties similar to those presented here (see also \cite{Abada:2008gs}). In the analysis of \cite{Bari_Riotto}, they explored the situation where the asymmetry is first generated by $N_2$ decays at a temperature where flavor effects \cite{flavor_effects} are important. Specifically, the relevant range of $10^9 \lesssim M_2 \lesssim 10^{12}$~GeV leads to a two-flavor regime where the lepton asymmetry will be stored in the $\tau$-component, as well as a coherent superposition of $(e,\mu)$-components. Subsequently, flavor dependent washout effects coming from interactions with $N_1$'s may not completely erase all components of the asymmetry generated by the $N_2$'s. One central conclusion in \cite{Bari_Riotto} is that, for this mechanism to generate enough asymmetry, the mass of the next-to-the-lightest RH neutrino must be about $M_2 \simeq 10^{11}$~GeV.

Inspecting the $M_2$-plot of Fig.~\ref{fig:mu_M123} (corresponding to the $\widehat{m}_f = \widehat{m}_u$ case), we can see that the condition of $M_2 \simeq 10^{11}$~GeV can be marginally met by a small region of the parameter space (near the various spikes in the region where $|m_1|$ is between $2\times 10^{-3}$ to $8\times 10^{-2}$~eV), whereas the $\widehat{m}_f = \widehat{m}_{d,e}$ cases are definitely ruled out for this scenario due to the smallness of $M_2$. Therefore, it appears that for some special values of $|m_1|$ with certain sets of phases $(\delta,\alpha_{1,2})$, leptogenesis via $N_2$ decays taking into account the effects of flavor is also possible (for the $\widehat{m}_f = \widehat{m}_u$ model) in addition to resonant leptogenesis.

Moreover, if this picture of flavored $N_2$-leptogenesis is indeed the mechansim responsible for generating the baryon asymmetry of the Universe, then the corresponding sets of low energy phases  in our model $(\delta,\alpha_{1,2})$ which make this possible will generally lead to modifications of the neutrinoless double beta decay rate through the quantity \cite{double_beta}
\begin{equation}
 m_{\beta\beta} \equiv \left|\sum_{i=1}^{3} U_{ei}^2\, m_i \right| \;.
\end{equation}
For example, taking $|m_1|= 0.070$~eV and assuming normal hierarchy, the phases implied by $N_2$-leptogenesis will lead to $m_{\beta\beta} \approx 0.047$~eV, which is a noticeable reduction from 0.070~eV in cases where both Majorana phases are turned off \footnote{The reason we have picked $|m_1|= 0.070$~eV in this discussion is because so far we have not found any set of phases for $|m_1|\gtrsim 0.08$ in which $N_2$-leptogenesis is actually viable.}. However, present experimental upper limits on $m_{\beta\beta}$ lie somewhere between 0.16 and 0.68~eV \cite{CUORICINO}, and so it is difficult to distinguish such differences. The detection of this may only be possible in future experiments such as CUORE \cite{CUORE} and GERDA \cite{GERDA} which have a projected sensitivity down to about 0.05~eV.

In summary, while the models presented in Sec.~\ref{Sec:models} do not generically lead to successful baryon asymmetry generation via thermal leptogenesis, some fine-tuned special cases do exist. It is possible that the enlargement of the workable parameter space for leptogenesis can result from modifications to the Higgs sector of these models, but such analyses are beyond the scope of this work.

\subsection{Collider Signatures}

It is interesting to note that in the model with $m_f = m_e$, the lightest heavy Majorana neutrino mass $M_1$ can be about 1 TeV making one wonder if it is possible to see signals of such a particle at the Large Hadron Collider (LHC) and or a future International Linear Collider (ILC). However, since the heavy Majorana neutrinos are dominantly right-handed singlets which do not have gauge interactions, the interactions of the heavy neutral leptons with SM gauge bosons arise through their mixing between light neutrinos.  The interaction Lagrangians are parameterized through mixing angles $V_{\ell N}$ ($\ell = e, \mu, \tau$) of order $m_\ell/M_i$ as per
\begin{eqnarray}
&&{\cal L}_W = -\frac{g}{\sqrt{2}}\, V_{\ell N}\, \overline{\ell}\, \gamma^\mu \, P_L\, N\, W_\mu + \text{h.c.}\;,\nonumber\\
&&{\cal L}_Z = -\frac{g}{2 \cos\theta_W}\, V_{\ell N}\, \overline{\nu}\, \gamma^\mu \, P_L\, N\, Z_\mu + \text{h.c.}\;,
\end{eqnarray}
where $P_{L,R} = (1 \mp \gamma_5)/2$.

With these interactions, it is possible to produce signals for heavy neutral leptons through
$q \overline{q}' \to W^{*} \to \ell N$ followed by $N \to \ell\, W$ or $\nu Z$. The production of $N$ by
$q \overline{q} \to Z^* \to \nu N$ is much harder to study due to large backgrounds. However, in a model-independent study in
Ref.~\cite{delAguila:2008cj}, such a mechanism was found to lead to a detectable heavy neutral lepton signal only if the mass is of 
order 100 GeV or less, for the initial stage of LHC running with
luminosity of order 10 $\text{fb}^{-1}$. 
Besides, the amplitudes of $V_{\ell N}$ in our models are far too small. Even assuming $|m_1| \simeq 0.2$~eV which will saturate the bound of (\ref{eq:nu_mass_limit}) and in the best case scenario with inverted hierarchy and special choice of phases, one obtains $|V_{eN}| \simeq 2.3\times 10^{-7}$ (with $M_1 \simeq 1.2\times 10^{3}$~GeV)  which is much less than the minimum \order{10^{-2}} required to produce an observable signal in any of the channels \cite{Aguila0502189}. The suppression is even greater for the $\mu$ or $\tau$ flavor.
As a result,
it is very difficult to detect the heavy neutral leptons through this mechanism even with an integrated luminosity up to 300 $\text{fb}^{-1}$.

If there is only one Higgs doublet, there is also a light neutrino and heavy neutral lepton interaction with the Higgs particle given by
\begin{eqnarray}
{\cal L}_H = -{g M_N\over 2 M_w}\,\left(V_{\ell N}\, \overline{\nu}\, P_R\, N\, H + \text{h.c.}\right)\;.
\end{eqnarray}
This interaction, although not of much help in the production of heavy neutral leptons through $q\bar q \to H^* \to \nu N$, 
does provide another channel for $N$ decay. If the Higgs mass is not too much larger than the W boson mass, the decay rate is similar to that for
$N \to \ell \,W$  or $\nu Z$.

In the models we are considering, there are several Higgs doublets. The neutral Higgs couplings to light neutrinos and heavy neutral leptons 
are then not necessarily proportional to $M_N V_{\ell N}$ and can increase the decay rate. Also, in our models there are
charged Higgs bosons interacting with light neutrinos and heavy neutral leptons which also provide additional channels for detection of the $N$'s. 
But 
given the smallness of the mixing $V_{\ell N}$ mentioned above,
it is still very difficult to detect a heavy neutral lepton with mass of order 1 TeV at the LHC even with 300 $\text{fb}^{-1}$ of luminosity.

Charged-Higgs couplings to charged-leptons and heavy neutral leptons may have interesting signals at an ILC through 
$e^+e^-\to H^+ H^-$ with t-channel heavy Higgs exchange, and $e^\pm e^\pm \to H^\pm H^\pm$ with u-channel $N$ exchange \cite{Atwood:2007zza}. In particular the processes  $e^\pm e^\pm \to H^\pm H^\pm$, are very sensitive the heavy neutral lepton mass. It has been shown in Ref.~\cite{Atwood:2007zza} that 
if $|V_{\ell N}|$ is in the range of $10^{-2}$ to $10^{-4}$, 
the ILC with an energy of 500 GeV can probe heavy neutral lepton masses up to $10^4$ TeV. In our case, the charged-Higgs coupling to charged-leptons and heavy neutral leptons can be larger than $V_{\ell N}\sim m_f/M_i$, but still too small to be probed using the processes mentioned above.

\section{Conclusion and Outlook}
\label{Sec:conclusion}

The main point of this paper was to demonstrate through general arguments backed up by explicit models that symmetries can be used to connect the RH Majorana neutrino mass matrix to low-energy observables such as charged-fermion masses, mixing angles and $CP$-violating phases.  If a model of this type were to actually describe nature, then the benefit would be that the high-mass seesaw sector would be completely determined from low-energy observations, improving the predictability and testability of the seesaw neutrino mass generation mechanism.  Since this mechanism can also be used to understand the cosmological matter-antimatter asymmetry through leptogenesis, such constrained models are also important for cosmology.

We focused on the simplest models of this type, which yielded $M_R \simeq \widehat{m}_{f}\, \Upmns^* \,\mnuhat^{-1}\, \Upmns^{\dagger}\, \widehat{m}_{f}$ where $f=e,d,u$.  Our phenomenological analysis showed that successful leptogenesis is possible for the $f=u$ case in certain fine-tuned corners of parameter space.  We also noted that the $e=f$ case can also supply a heavy neutral lepton with a mass of about 1 TeV, opening the prospect for collider detection, though detailed analysis showed that this mass is still too large to plausibly expect detection at either the LHC or a future ILC.

Future work in this are could explore a possible role for the CKM matrix rather than the more obvious PMNS matrix in the formula for $M_R$.  Also, the use of Clebsch-Gordan coefficients to generalize the relationship between the neutrino Dirac mass matrix and $\widehat{m}_f$ away from being a strict equality is another obvious line of investigation.  Finally, our explicit models used flavor symmetry to render the right-handed diagonalization matrices to be simply identity matrices.  It could also be of interest to loosen this constraint.

\section*{Acknowledgements}
SSCL and RRV thank Nicole Bell for discussions. XGH would like to thank KITPC for hospitality where part of this work was done. This work was supported in part by the Australian Research Council, the Commonwealth of Australia, the NSC and NCTS.

\appendix

\section{properties of the $A_4$ group}\label{app:A4}

$A_4$ is the alternating group of order 4. It is isomorphic to the group representing the proper rotational symmetries of a regular tetrahedron. It has 12 elements and 4 conjugacy classes: one set containing the identity, two sets containing four 3-fold rotations each and one set of three 2-fold rotations. By the dimensionality theorem, we know that $A_4$ must have four irreducible representations: $\rep, \repp,\reppp$ and $\rept$, where $\rep$ is the trivial representation, $\repp$ and $\reppp$ are non-trivial one-dimensional representations that are complex conjugate of each other, while $\rept$ is a real three-dimensional representation.

Some basic tensor product rules:
\begin{align}
 &\rep \otimes \rep = \rep \;,\\
 &\repp \otimes \reppp = \rep \;,\\
 &\repp \otimes \repp = \reppp \;,\\
 &\rept \otimes \rept = \rep \oplus \repp \oplus \reppp \oplus \rept_a \oplus \rept_s\;,\label{app:3x3}
\end{align}
where subscripts $a$ and $s$ denote ``asymmetric'' and ``symmetric'' respectively. Suppose $x_{\rept}=(x_1,x_2,x_3)$ and $y_{\rept}=(y_1,y_2,y_3)$ are triplets in $A_4$. Then Eq.~(\ref{app:3x3}) means 
\begin{align}
 (x_{\rept}\; y_{\rept})_{\rep} &= x_1 y_1 + x_2 y_2 + x_3 y_3\;,\\
 (x_{\rept}\; y_{\rept})_{\repp} &= x_1 y_1 + \omega x_2 y_2 + \omega^2 x_3 y_3\;,\\
 (x_{\rept}\; y_{\rept})_{\reppp} &= x_1 y_1 + \omega^2 x_2 y_2 + \omega x_3 y_3\;,\\
 (x_{\rept}\; y_{\rept})_{\rept_a} &= (x_2 y_3 - x_3 y_2\;, x_3 y_1 - x_1 y_3\;, x_1 y_2 - x_2 y_1)\;,\\
 (x_{\rept}\; y_{\rept})_{\rept_s} &= (x_2 y_3 + x_3 y_2\;, x_3 y_1 + x_1 y_3\;, x_1 y_2 + x_2 y_1)\;,
\end{align}
where $\omega = e^{2\pi i/3}$ and we have abbreviated $(x_{\rept}\otimes y_{\rept})$ with $(x_{\rept}\; y_{\rept})$.


\begin{thebibliography}{99}


\bibitem{neutrinos_exp}
  B.~T.~Cleveland {\it et al.},
  Nucl.\ Phys.\ Proc.\ Suppl.\  {\bf 38}, 47 (1995);
  Y.~Fukuda {\it et al.}  [Super-Kamiokande Collaboration],
  Phys.\ Rev.\ Lett.\  {\bf 82}, 2430 (1999)
  [arXiv:hep-ex/9812011];
   Y.~Fukuda {\it et al.}  [Super-Kamiokande Collaboration],
  Phys.\ Rev.\ Lett.\  {\bf 81}, 1562 (1998)
  [arXiv:hep-ex/9807003];
  K.~Lande {\it et al.},
  Nucl.\ Phys.\ Proc.\ Suppl.\  {\bf 77}, 13 (1999);
  D.~N.~Abdurashitov {\it et al.}  [SAGE Collaboration],
  Nucl.\ Phys.\ Proc.\ Suppl.\  {\bf 77}, 20 (1999);
  T.~A.~Kirsten  [GALLEX and GNO Collaborations],
  Nucl.\ Phys.\ Proc.\ Suppl.\  {\bf 77}, 26 (1999);
  Y.~Fukuda {\it et al.}  [Super-Kamiokande Collaboration],
  Phys.\ Lett.\ B {\bf 436}, 33 (1998)
  [arXiv:hep-ex/9805006];
  C.~Athanassopoulos {\it et al.}  [LSND Collaboration],
  Phys.\ Rev.\ C {\bf 54}, 2685 (1996)
  [arXiv:nucl-ex/9605001];
  C.~Athanassopoulos {\it et al.}  [LSND Collaboration],
  Phys.\ Rev.\ C {\bf 58}, 2489 (1998)
  [arXiv:nucl-ex/9706006];
  Q.~R.~Ahmad {\it et al.}  [SNO Collaboration],
  Phys.\ Rev.\ Lett.\  {\bf 89}, 011301 (2002)
  [arXiv:nucl-ex/0204008];
  S.~N.~Ahmed {\it et al.}  [SNO Collaboration],
  Phys.\ Rev.\ Lett.\  {\bf 92}, 181301 (2004)
  [arXiv:nucl-ex/0309004];
  K.~Eguchi {\it et al.}  [KamLAND Collaboration],
  Phys.\ Rev.\ Lett.\  {\bf 90}, 021802 (2003)
  [arXiv:hep-ex/0212021].

\bibitem{type1_seesaw}
  P.~Minkowski,
  Phys.\ Lett.\ B {\bf 67}, 421 (1977);
  T.~Yanagida, in {\it Workshop on Unified Theories}, KEK report 79-18 p.95 (1979);
  M.~Gell-Mann, P.~Ramond, R.~Slansky, 
  in {\it Supergravity} (North Holland, Amsterdam, 1979)
  eds. P.~van~Nieuwenhuizen, D.~Freedman, p.315;
  S.~L.~Glashow, in {\it 1979 Cargese Summer Institute on Quarks and Leptons} (Plenum Press,
  New York, 1980) eds. M.~Levy, J.-L.~Basdevant, D.~Speiser, J.~Weyers, R.~Gastmans and M.~Jacobs,
  p.687; 
  R.~Barbieri, D.~V.~Nanopoulos, G.~Morchio and F.~Strocchi,
  Phys.\ Lett.\ B {\bf 90}, 91 (1980);
  R.~N.~Mohapatra and G.~Senjanovic,
  Phys.\ Rev.\ Lett.\  {\bf 44}, 912 (1980);
  G.~Lazarides, Q.~Shafi and C.~Wetterich,
  Nucl.\ Phys.\  B {\bf 181}, 287 (1981).

\bibitem{type2_seesaw}
  W.~Konetschny and W.~Kummer,
  Phys.\ Lett.\  B {\bf 70}, 433 (1977);
%
 T.~P.~Cheng and L.~F.~Li,
  Phys.\ Rev.\  D {\bf 22}, 2860 (1980);
%
 G.~Lazarides, Q.~Shafi and C.~Wetterich,
  Nucl.\ Phys.\  B {\bf 181}, 287 (1981);
%
 J.~Schechter and J.~W.~F.~Valle,
  Phys.\ Rev.\  D {\bf 22}, 2227 (1980);
%
 R.~N.~Mohapatra and G.~Senjanovic,
  Phys.\ Rev.\  D {\bf 23}, 165 (1981).

\bibitem{type3_seesaw}
  R.~Foot, H.~Lew, X.~G.~He and G.~C.~Joshi,
  Z.\ Phys.\  C {\bf 44}, 441 (1989).


\bibitem{zeebabu}
  A.~Zee,
  Nucl.\ Phys.\  B {\bf 264}, 99 (1986);
%
  K.~S.~Babu,
  Phys.\ Lett.\  B {\bf 203}, 132 (1988).


\bibitem{Fukugita:1986hr}
  M.~Fukugita and T.~Yanagida,
  Phys.\ Lett.\ B {\bf 174}, 45 (1986).
    
\bibitem{BDP}
See for example:
  W.~Buchm\"uller, P.~Di Bari and M.~Pl\"umacher,
  Annals Phys.\  {\bf 315}, 305 (2005)
  [arXiv:hep-ph/0401240],
%
  W.~Buchmuller, R.~D.~Peccei and T.~Yanagida,
  Ann.\ Rev.\ Nucl.\ Part.\ Sci.\  {\bf 55}, 311 (2005)
  [arXiv:hep-ph/0502169].
  
\bibitem{ARS}
  E.~K.~Akhmedov, V.~A.~Rubakov and A.~Y.~Smirnov,
  Phys.\ Rev.\ Lett.\  {\bf 81}, 1359 (1998)
  [arXiv:hep-ph/9803255].


\bibitem{WDM}
 See for example:   
  K.~Abazajian, G.~M.~Fuller and M.~Patel,
  Phys.\ Rev.\  D {\bf 64}, 023501 (2001)
  [arXiv:astro-ph/0101524].
%
  T.~Asaka, M.~Shaposhnikov and A.~Kusenko,
  Phys.\ Lett.\  B {\bf 638}, 401 (2006)
  [arXiv:hep-ph/0602150],
%
  T.~Asaka, S.~Blanchet and M.~Shaposhnikov,
  Phys.\ Lett.\  B {\bf 631}, 151 (2005)
  [arXiv:hep-ph/0503065],
and references within.
 
 
\bibitem{mixingdata}
  S.~N.~Ahmed {\it et al.}  [SNO Collaboration],
  Phys.\ Rev.\ Lett.\  {\bf 92}, 181301 (2004) 
  [arXiv:nucl-ex/0309004];
%
  M.~H.~Ahn {\it et al.}  [K2K Collaboration],
  Phys.\ Rev.\ Lett.\  {\bf 90}, 041801 (2003) 
  [arXiv:hep-ex/0212007];
  G.~L.~Fogli, E.~Lisi, A.~Marrone, A.~Palazzo and D.~Montanino,
  Nucl.\ Phys.\ Proc.\ Suppl.\  {\bf 118}, 177 (2003)
  [arXiv:hep-ph/0310012];
%
  G.~L.~Fogli, E.~Lisi, A.~Marrone, A.~Melchiorri, A.~Palazzo, P.~Serra and J.~Silk,
  Phys.\ Rev.\ D {\bf 70}, 113003 (2004)
  [arXiv:hep-ph/0408045] (and references therein);
  W.~M.~Yao {\it et al.}  [Particle Data Group],
  J.\ Phys.\ G {\bf 33}, 1 (2006).
  
\bibitem{Schwetz:2008er}
  T.~Schwetz, M.~Tortola and J.~W.~F.~Valle,
  arXiv:0808.2016 [hep-ph].

\bibitem{Berezhiani:1990jj}
  Z.~G.~Berezhiani and M.~Y.~Khlopov,
  Sov.\ J.\ Nucl.\ Phys.\  {\bf 51}, 935 (1990)
  [Yad.\ Fiz.\  {\bf 51}, 1479 (1990)].

  
\bibitem{Georgi:1974sy}
  H.~Georgi and S.~L.~Glashow,
  Phys.\ Rev.\ Lett.\  {\bf 32}, 438 (1974).
  
\bibitem{Barr:1981qv}
  S.~M.~Barr,
  Phys.\ Lett.\  B {\bf 112}, 219 (1982).
  
\bibitem{LR_Mohapatra.Pati}
  R.~N.~Mohapatra and J.~C.~Pati,
  Phys.\ Rev.\  D {\bf 11}, 2558 (1975),
  %
  R.~N.~Mohapatra and J.~C.~Pati,
  Phys.\ Rev.\  D {\bf 11}, 566 (1975).

\bibitem{Pati:1974yy}
  J.~C.~Pati and A.~Salam,
  Phys.\ Rev.\  D {\bf 10}, 275 (1974)
  [Erratum-ibid.\  D {\bf 11}, 703 (1975)].
  
  
\bibitem{Volkas:1995yn}
  R.~R.~Volkas,
  Phys.\ Rev.\  D {\bf 53}, 2681 (1996)
  [arXiv:hep-ph/9507215].


\bibitem{QL_symmodel}
  R.~Foot and H.~Lew,
  Phys.\ Rev.\  D {\bf 41}, 3502 (1990),
%
  R.~Foot, H.~Lew and R.~R.~Volkas,
  Phys.\ Rev.\  D {\bf 44}, 1531 (1991).

\bibitem{Georgi:1979df}
  H.~Georgi and C.~Jarlskog,
  Phys.\ Lett.\  B {\bf 86}, 297 (1979).

\bibitem{Low:2003dz}
  C.~I.~Low and R.~R.~Volkas,
  Phys.\ Rev.\  D {\bf 68}, 033007 (2003)
  [arXiv:hep-ph/0305243].


\bibitem{TBmixing}
  P.~F.~Harrison, D.~H.~Perkins and W.~G.~Scott,
  Phys.\ Lett.\  B {\bf 530}, 167 (2002)
  [arXiv:hep-ph/0202074],
%
  P.~F.~Harrison and W.~G.~Scott,
  Phys.\ Lett.\  B {\bf 535}, 163 (2002)
  [arXiv:hep-ph/0203209];
  Phys.\ Lett.\  B {\bf 557}, 76 (2003)
  [arXiv:hep-ph/0302025],
%
  Z.~Z.~Xing,
  Phys.\ Lett.\  B {\bf 533}, 85 (2002)
  [arXiv:hep-ph/0204049],
%
  X.~G.~He and A.~Zee,
  Phys.\ Lett.\  B {\bf 560}, 87 (2003)
  [arXiv:hep-ph/0301092].
  
\bibitem{A4_ref}
  E.~Ma and G.~Rajasekaran,
  Phys.\ Rev.\  D {\bf 64}, 113012 (2001)
  [arXiv:hep-ph/0106291],
%
  K.~S.~Babu, E.~Ma and J.~W.~F.~Valle,
  Phys.\ Lett.\  B {\bf 552}, 207 (2003)
  [arXiv:hep-ph/0206292],
  %
  E.~Ma,
  Phys.\ Rev.\  D {\bf 70}, 031901 (2004)
  [arXiv:hep-ph/0404199],
  %
  E.~Ma,
  arXiv:hep-ph/0409075,
  %
  G.~Altarelli and F.~Feruglio,
  Nucl.\ Phys.\  B {\bf 720}, 64 (2005)
  [arXiv:hep-ph/0504165],
  %
  G.~Altarelli and F.~Feruglio,
  Nucl.\ Phys.\  B {\bf 741}, 215 (2006)
  [arXiv:hep-ph/0512103],
  %
  K.~S.~Babu and X.~G.~He,
  arXiv:hep-ph/0507217,
  %
  A.~Zee,
  Phys.\ Lett.\  B {\bf 630}, 58 (2005)
  [arXiv:hep-ph/0508278],
  %
  E.~Ma,
  Phys.\ Rev.\  D {\bf 72}, 037301 (2005)
  [arXiv:hep-ph/0505209],
  %
  E.~Ma,
  Mod.\ Phys.\ Lett.\  A {\bf 20}, 2601 (2005)
  [arXiv:hep-ph/0508099].
  
\bibitem{Lew:1992rr}
  H.~Lew and R.~R.~Volkas,
  Phys.\ Rev.\  D {\bf 47}, 1356 (1993)
  [arXiv:hep-ph/9209288].

\bibitem{He:2006dk}
  X.~G.~He, Y.~Y.~Keum and R.~R.~Volkas,
  JHEP {\bf 0604}, 039 (2006)
  [arXiv:hep-ph/0601001].
  
  
\bibitem{WMAP}
  E.~Komatsu {\it et al.}  [WMAP Collaboration],
  arXiv:0803.0547 [astro-ph].
 
\bibitem{Akhmedov03}
   E.~K.~Akhmedov, M.~Frigerio and A.~Y.~Smirnov,
  JHEP {\bf 0309}, 021 (2003)
  [arXiv:hep-ph/0305322].

\bibitem{Branco02} 
  G.~C.~Branco, R.~Gonzalez Felipe, F.~R.~Joaquim and M.~N.~Rebelo,
  Nucl.\ Phys.\  B {\bf 640}, 202 (2002)
  [arXiv:hep-ph/0202030].

\bibitem{running_mass}
  H.~Fusaoka and Y.~Koide,
  Phys.\ Rev.\ D {\bf 57}, 3986 (1998)
  [arXiv:hep-ph/9712201].


\bibitem{Nezri00}
  E.~Nezri and J.~Orloff,
  JHEP {\bf 0304}, 020 (2003)
  [arXiv:hep-ph/0004227].



\bibitem{resonant}
  M.~Flanz, E.~A.~Paschos and U.~Sarkar,
  Phys.\ Lett.\  B {\bf 345}, 248 (1995)
  [Erratum-ibid.\  B {\bf 382}, 447 (1996)]
  [arXiv:hep-ph/9411366];
%
 M.~Flanz, E.~A.~Paschos, U.~Sarkar and J.~Weiss,
  Phys.\ Lett.\  B {\bf 389}, 693 (1996)
  [arXiv:hep-ph/9607310];
%
  A.~Pilaftsis,
  Phys.\ Rev.\  D {\bf 56}, 5431 (1997)
  [arXiv:hep-ph/9707235].
%
 See also
%
 A.~Pilaftsis and T.~E.~J.~Underwood,
  Nucl.\ Phys.\  B {\bf 692}, 303 (2004)
  [arXiv:hep-ph/0309342];
%
 A.~Pilaftsis and T.~E.~J.~Underwood,
  Phys.\ Rev.\  D {\bf 72}, 113001 (2005)
  [arXiv:hep-ph/0506107];
%
 A.~De Simone and A.~Riotto,
  JCAP {\bf 0708}, 013 (2007)
  [arXiv:0705.2183 [hep-ph]]
and references therein.


\bibitem{DiBari:2005st}
  P.~Di Bari,
  Nucl.\ Phys.\ B {\bf 727}, 318 (2005) 
  [arXiv:hep-ph/0502082].

\bibitem{Bari_Riotto}
  P.~Di Bari and A.~Riotto,
  arXiv:0809.2285 [hep-ph].

\bibitem{Abada:2008gs}
  A.~Abada, P.~Hosteins, F.~X.~Josse-Michaux and S.~Lavignac,
  arXiv:0808.2058 [hep-ph].


\bibitem{flavor_effects}
  R.~Barbieri, P.~Creminelli, A.~Strumia and N.~Tetradis,
  Nucl.\ Phys.\ B {\bf 575}, 61 (2000)
  [arXiv:hep-ph/9911315],
  %
  O.~Vives,
  Phys.\ Rev.\ D {\bf 73}, 073006 (2006)
  [arXiv:hep-ph/0512160],
  %
  E.~Nardi, Y.~Nir, E.~Roulet and J.~Racker,
  JHEP {\bf 0601}, 164 (2006)
  [arXiv:hep-ph/0601084],
  %
  A.~Abada, S.~Davidson, F.~X.~Josse-Michaux, M.~Losada and A.~Riotto,
  JCAP {\bf 0604}, 004 (2006)
  [arXiv:hep-ph/0601083],
  A.~Abada, S.~Davidson, A.~Ibarra, F.~X.~Josse-Michaux, M.~Losada and A.~Riotto,
  JHEP {\bf 0609}, 010 (2006)
  [arXiv:hep-ph/0605281],
  %
  S.~Blanchet and P.~Di Bari,
  JCAP {\bf 0703}, 018 (2007)
  [arXiv:hep-ph/0607330].


\bibitem{double_beta}
  S.~M.~Bilenky and S.~T.~Petcov,
  Rev.\ Mod.\ Phys.\  {\bf 59}, 671 (1987)
  [Erratum-ibid.\  {\bf 61}, 169.1989\ ERRAT,60,575 (1989\ ERRAT,60,575-575.1988)];
 %
  S.~R.~Elliott and P.~Vogel,
  Ann.\ Rev.\ Nucl.\ Part.\ Sci.\  {\bf 52}, 115 (2002)
  [arXiv:hep-ph/0202264].
  
\bibitem{CUORICINO}
  C.~Arnaboldi {\it et al.}  [CUORICINO Collaboration],
  Phys.\ Rev.\  C {\bf 78}, 035502 (2008)
  [arXiv:0802.3439 [hep-ex]].
  
\bibitem{CUORE}
  C.~Arnaboldi {\it et al.}  [CUORE Collaboration],
  Nucl.\ Instrum.\ Meth.\  A {\bf 518}, 775 (2004)
  [arXiv:hep-ex/0212053];
%
  E.~Fiorini,
  Phys.\ Rept.\  {\bf 307}, 309 (1998);
%
  R.~Ardito {\it et al.},
  arXiv:hep-ex/0501010.


\bibitem{GERDA}
  S.~Schonert {\it et al.}  [GERDA Collaboration],
  Nucl.\ Phys.\ Proc.\ Suppl.\  {\bf 145}, 242 (2005);
%
  I.~Abt {\it et al.},
  arXiv:hep-ex/0404039.
  



\bibitem{delAguila:2008cj}
  F.~del Aguila and J.~A.~Aguilar-Saavedra,
  arXiv:0808.2468 [hep-ph].


\bibitem{Aguila0502189}
  F.~del Aguila, J.~A.~Aguilar-Saavedra, A.~Martinez de la Ossa and D.~Meloni,
  Phys.\ Lett.\  B {\bf 613}, 170 (2005)
  [arXiv:hep-ph/0502189].


\bibitem{Atwood:2007zza}
  D.~Atwood, S.~Bar-Shalom and A.~Soni,
  Phys.\ Rev.\  D {\bf 76}, 033004 (2007)
  [arXiv:hep-ph/0701005].


 
\end{thebibliography}
\end{document}